\documentclass[reprint,superscriptaddress,nofootinbib,english,preprintnumbers,amsmath,amssymb,aps,prx]{revtex4-2}
\usepackage{graphicx}
\usepackage{dsfont}
\usepackage{xcolor}
\usepackage{multirow}
\usepackage{breakurl}
\usepackage[normalem]{ulem}
\usepackage{enumitem} 
\usepackage{slashed}
\usepackage{charter} 
\usepackage[charter]{mathdesign}

\allowdisplaybreaks

\definecolor{refcol}{RGB}{178,34,34}%{178,34,34}%{0,100,170}%{0,0,205}
\usepackage{microtype}
\usepackage[colorlinks,linkcolor=refcol,citecolor=refcol,urlcolor=refcol]{hyperref}

\graphicspath{{./figures/}}

\newcommand{\tr}{\mathrm{tr}}
\def\det{{\mathrm{det}}}

\def\eq#1{(\ref{#1})}

\def\Eq#1{Eq.~(\ref{#1})}

\def\Fig#1{Fig.~\ref{#1}}

\def\Sec#1{Sec.~\ref{#1}}

\definecolor{blue}{rgb}{0,0,1}

\definecolor{green}{rgb}{0,1,0}

\definecolor{red}{rgb}{1,0,0}

\begin{document}

\title{Medium induced mixing, spatial modulations and critical modes in QCD}

\author{Maximilian Haensch}
\affiliation{Institute for Theoretical Physics, Justus Liebig University Giessen, Heinrich-Buff-Ring 16, 35392 Giessen, Germany}

\author{Fabian Rennecke}
\email{fabian.rennecke@theo.physik.uni-giessen.de} 
\affiliation{Institute for Theoretical Physics, Justus Liebig University Giessen, Heinrich-Buff-Ring 16, 35392 Giessen, Germany}
\affiliation{Helmholtz Research Academy Hesse for FAIR (HFHF), Campus Giessen, Giessen, Germany}

\author{Lorenz von Smekal}
\affiliation{Institute for Theoretical Physics, Justus Liebig University Giessen, Heinrich-Buff-Ring 16, 35392 Giessen, Germany}
\affiliation{Helmholtz Research Academy Hesse for FAIR (HFHF), Campus Giessen, Giessen, Germany}

\begin{abstract}
The mixing between the chiral condensate and the density in hot and dense QCD matter is familiar. We show that the mixing relevant for the ground state is considerably more extensive, and in particular also involves gluonic degrees of freedom. As a result, the Hessian of the QCD effective action is non-Hermitian, but retains a symmetry under combined charge- and complex conjugation. This can lead to complex-conjugate pairs of eigenvalues of this Hessian, signaling regimes with spatially modulated correlations. Furthermore, based on the analytic structure of the quark determinant at a chiral critical point, we demonstrate that the corresponding massless critical mode is composed of the chiral condensate, the density and the Polyakov loops. Due to an avoided crossing, the critical mode turns out to be disconnected from the chiral condensate in vacuum. We present general arguments for all these features and illustrate them through explicit model calculations.
\end{abstract}

\maketitle

%\tableofcontents

%%%%%%%%%%%%%%%%%%%%%%%%%%%%%%%%%%%%%%%%%%%%%%%%%%
%%%%%%%%%%%%%%%%%%%%%%%%%%%%%%%%%%%%%%%%%%%%%%%%%%
\section{Introduction}\label{sec:intro}
%%%%%%%%%%%%%%%%%%%%%%%%%%%%%%%%%%%%%%%%%%%%%%%%%%
%%%%%%%%%%%%%%%%%%%%%%%%%%%%%%%%%%%%%%%%%%%%%%%%%%
It is by now well established that QCD features a chiral crossover at small baryon chemical potential \cite{Aoki:2006we, Borsanyi:2010bp, HotQCD:2018pds}. Model studies suggest that this crossover might turn into a critical endpoint (CEP) at larger chemical potential, where the transition is of second order \cite{Stephanov:2004wx, Fukushima:2010bq}. This expectation is supported by recent results of functional continuum methods, which have established that the chiral crossover becomes steeper as the chemical potential is increased \cite{Fischer:2018sdj, Fu:2019hdw, Gao:2020fbl, Gunkel:2021oya,Fu:2022gou}. Assuming that no other phases emerge, these works indeed predict a CEP roughly around temperatures $T\approx 100$\, MeV and baryon chemical potentials $\mu_B \approx 600$\,MeV. 

However, most studies that find a CEP assume a homogeneous ground state. Model studies that do not make this assumption tend to find inhomogeneous or crystalline phases with spatial modulations \cite{Fukushima:2010bq,Buballa:2014tba}. A precursor for such phases, a moat regime where the particle energy is minimized at nonzero momentum \cite{Pisarski:2020dnx, Pisarski:2021qof, Rennecke:2021ovl}, has recently been discovered in QCD in Ref.\ \cite{Fu:2019hdw} in the vicinity of the CEP. This is a strong indication that, for example, either an inhomogeneous phase \cite{Buballa:2014tba}, a liquid crystal \cite{Lee:2015bva, Hidaka:2015xza} or a quantum pion liquid \cite{Pisarski:2020dnx} could exists at large density. The latter two possibilities arise from fluctuation-induced instabilities. In any case, these phases could potentially wash-out the CEP, replacing it with a Lifshitz point, or a Lifshitz regime \cite{Pisarski:2018bct}. This is also relevant for heavy-ion collisions, where a moat regime might leave observable signatures \cite{Pisarski:2021qof, Rennecke:2023xhc, Fukushima:2023tpv}.

In order to understand the phase structure, the relevant degrees of freedom need to be identified correctly. To this end, it is important to take into account any mixing that affects the ground state properties of the system. For example, the ground state of nuclear matter is characterized, among other things, by a non-vanishing chiral condensate $\langle \bar\psi \psi \rangle$ and the quark/baryon number density $\langle \bar\psi \gamma^0 \psi \rangle$. These can be identified with the condensates of "mesons" in the isoscalar-scalar channel and the temporal component in the isoscalar-vector channel, $\sigma$ and $\omega^0$. It is well known that the chiral condensate and the density mix in dense QCD matter, see, e.g., Refs.\ \cite{Lim:1989ma, Kunihiro:1991qu}. Mixing means that there is a nonzero linear coupling $\sim\sigma \omega^0$. Hence, the relevant degree of freedom characterizing the chiral properties of the system is not the $\sigma$-meson itself, but a mixture of $\sigma$ and $\omega^0$.

This has far-reaching consequences. The potential existence of a CEP has motivated experimental searches with ultrarelativistic heavy-ion collisions \cite{Luo:2017faz, Bzdak:2019pkr, Almaalol:2022xwv}. Many of these searches rely on the sensitivity of fluctuations in net-baryon multiplicities on the critical physics in the vicinity of the CEP \cite{Stephanov:1999zu, Hatta:2003wn, Stephanov:2008qz, Stephanov:2011pb}. A non-monotonic beam-energy dependence has been proposed as a signature of a CEP in these works. Although it is now known that this is not a smoking-gun since this behavior can also arise from a steep crossover \cite{Fu:2021oaw},
there is little doubt that a system created in a heavy-ion collision is affected by critical physics if it evolves close enough to a CEP. Most notably, critical long-range correlations lead to equilibration times that exceed the timescales related to the expansion and cooling of matter in a heavy-ion collision \cite{Rajagopal:1992qz,Berdnikov:1999ph}. Due to this critical slowing down, the system will fall out of equilibrium. In addition to the symmetries and the dimensionality of the system which determine the static universal properties, additional conserved charges and hydrodynamic modes need to be taken into account in order to describe the resulting dynamic universal behavior near the CEP \cite{Hohenberg:1977ym, Berdnikov:1999ph, Son:2004iv}. This is affected by mixing.

Without mixing, the CEP is signaled by a massless spacelike mode in the $\sigma$ channel \cite{Fujii:2003bz}. Since the order parameter is not conserved, this leads to so-called Model A dynamics \cite{Hohenberg:1977ym}. Taking into account the dynamics of the coupled dissipative and diffusive fluctuations of the mixture between the chiral order parameter and conserved baryon density in an otherwise static medium, it has been shown that only one hydrodynamic mode is provided by the slow diffusive fluctuations of the latter \cite{Son:2004iv}, leading to the dynamic universality class of Model B \cite{Hohenberg:1977ym}. If the motion of the medium is considered as well, coupling this mode to the energy-momentum tensor, from the potentially four additional hydrodynamic modes, only the two transverse components of the conserved momentum density lead to additional slow  modes. Non-trivial reversible mode-mode couplings between these hydrodynamic modes then put the QCD CEP in the dynamic universality class of Model H \cite{Son:2004iv}, which is that of a liquid-gas transition in a pure fluid \cite{Hohenberg:1977ym}.

Furthermore, it has been demonstrated in Ref.\ \cite{Nishimura:2016yue} that the combination of $\sigma$-$\omega^0$ mixing and the short-distance repulsion of nucleons induced by $\omega^0$-exchange leads to a non-Hermitian meson mass matrix at finite density, where the symmetry under charge conjugation, $\mathcal{C}$, is broken. However, QCD retains a symmetry under $\mathcal{C}\mathcal{K}$ transformations, where $\mathcal{K}$ is complex conjugation, and the eigenvalues of the Mass matrix can come in complex-conjugate pairs. If this happens, correlations show spatial modulations, much like in a moat regime. The $\mathcal{CK}$-symmetry of finite-density QCD has been argued to be a form of $\mathcal{P}\mathcal{T}$, and the implications for the phase structure of various systems has been discussed in Refs.\ \cite{Nishimura:2014rxa, Nishimura:2014kla, Nishimura:2015lit, Nishimura:2016yue, Schindler:2019ugo, Schindler:2021otf}.

Since mixing clearly plays an important role for QCD matter, a detailed and systematic understanding of the effects relevant for the phase structure is important. To this end, we present a general strategy to study mixing in dense QCD matter based on the functional quark determinant of the path integral. The aforementioned $\sigma$-$\omega^0$-mixing naturally arises from this approach at finite density away from the chiral limit. However, the underlying mechanism is more general, and in particular applies whenever a field or composite operator couples to the quarks in a way similar to the chemical potential. Interestingly, this is also the case for the Polyakov loop $\Phi$ and its conjugate $\bar \Phi$, which is used as an order parameter for confinement in the pure gauge theory and serves as a measure for nontrivial electric gluon correlations in finite temperature QCD as well \cite{Fukushima:2017csk}. The confined ground state of nuclear matter at finite density is hence also characterized by nontrivial $\Phi$ and $\bar\Phi$. Their mixing with the chiral condensate has been noted previously \cite{Fukushima:2003fw}, but its consequences have not been explored yet.

We systematically investigate the mixing induced by fluctuations around the ground state of nuclear matter, which is characterized by the chiral condensate, the baryon number density and the Polyakov loops. We explore the physical implications based on both, general arguments and explicit model calculations. From the discussion above we expect the resulting Hessian to be non-Hermitian, facilitating in particular the existence of spatially modulated regimes in the phase diagram. Furthermore, we employ arguments based on the analytic structure of the free energy at the CEP to clarify the nature of the critical mode.
Possible implications on the critical dynamics are beyond the scope of our present work. 

This paper is organized as follows. In Secs.\ \ref{sec:mix} and \ref{sec:hess} we derive the medium-induced mixing in QCD and the resulting Hessian using a saddle-point expansion of the quark determinant. We then discuss the possible implications for the critical mode and the phase structure in Secs.\ \ref{sec:crit} and \ref{sec:mod}. To illustrate our general arguments, we study different Polyakov--Quark-Meson models, which effectively capture basic features of the QCD phase transition, in \Sec{sec:ex}. In \Sec{sec:PQM} we study the basic model with mean-fields for the chiral condensate and the Polyakov loops, and in \Sec{sec:PQMV} we add  vector repulsion and vacuum fluctuations. We present a summary of our findings and a brief outlook in \Sec{sec:conc}.

%%%%%%%%%%%%%%%%%%%%%%%%%%%%%%%%%%%%%%%%%%%%%%%%%%
%%%%%%%%%%%%%%%%%%%%%%%%%%%%%%%%%%%%%%%%%%%%%%%%%%
\section{Induced mixing}\label{sec:crimo}
%%%%%%%%%%%%%%%%%%%%%%%%%%%%%%%%%%%%%%%%%%%%%%%%%%
%%%%%%%%%%%%%%%%%%%%%%%%%%%%%%%%%%%%%%%%%%%%%%%%%%

%%%%%%%%%%%%%%%%%%%%%%%%%%%%%%%%%%%%%%%%%%%%%%%%%%
%%%%%%%%%%%%%%%%%%%%%%%%%%%%%%%%%%%%%%%%%%%%%%%%%%
\subsection{Saddle-point approximation}\label{sec:mix}
%%%%%%%%%%%%%%%%%%%%%%%%%%%%%%%%%%%%%%%%%%%%%%%%%%
%%%%%%%%%%%%%%%%%%%%%%%%%%%%%%%%%%%%%%%%%%%%%%%%%%

We start by discussing how mixing through linear couplings in QCD at finite chemical potential arises. Our focus is on the relevant ground-state properties of dense nuclear matter. To this end, we note that fundamental interactions in QCD give rise to a Yukawa-type interaction between quarks and $\omega$ mesons, $\bar\psi \gamma^\mu \omega_\mu \psi$. This interaction arises from the resonance of a four-quark interaction in the isoscalar vector channel, $(\bar\psi \gamma^\mu\psi)^2$ and can be derived from first principles using, e.g., dynamical hadronization \cite{Gies:2001nw, Pawlowski:2005xe, Braun:2009ewx, Floerchinger:2009uf, Mitter:2014wpa, Braun:2014ata,Rennecke:2015eba,Cyrol:2017ewj,Fu:2019hdw,Fukushima:2021ctq}. For the present argument, we focus on the coupling between the quarks and $\omega_0$,
\begin{align}
i h_\omega \bar\psi \gamma^0 \omega_0 \psi\,,
\end{align}
where $h_\omega$ is the relevant Yukawa coupling. Although this restriction is not necessary for the following arguments, we assume a repulsive vector interaction, so $i h_\omega$ has to be imaginary. This assumption is supported by nuclear matter phenomenology \cite{Serot:1997xg}.
$\omega_0$ couples to the quark density in the same way as the quark chemical potential. Its equations of motion (EoM) can have nontrivial solutions in QCD, i.e.\ $\omega_0$ can acquire a vacuum expectation value (VEV). This VEV is necessarily related to the quark density $n_q$, so that 
\begin{align}\label{eq:baromega}
-i\bar\omega_0 = \langle \omega_0\rangle \sim n_q\,. 
\end{align}
We note that $\bar\omega_0$ is purely imaginary.
Furthermore, we introduce a scalar field, $\sigma$, whose VEV is directly related to the chiral condensate,
\begin{align}
\bar\sigma \sim \langle\bar\psi \psi\rangle\,.
\end{align}
It couples to the quarks through a Yukawa interaction,
\begin{align}
h_\sigma \bar\psi\sigma\psi\,.
\end{align}
Chiral symmetry dictates that other fields, in particular pseudo-scalars, couple to the quarks in a similar way. But the details are irrelevant for the present discussion.
The Dirac operator at finite chemical potential has the following form then:
\begin{align}
\mathcal{M} = \gamma^\mu D_\mu + m + h_\sigma\sigma +\gamma^0(\mu + ih_\omega\omega_0)\,,
\end{align}
where $m$ is the current quark mass and the covariant derivative in the fundamental representation is
\begin{align}
D_\mu = \partial_\mu - i g A_\mu\,,
\end{align}
with $A_\mu = A_\mu^aT_a$ with the $SU(N_c)$ generators $T_a$, $a = 1,\dots,N_c^2-1$.
For now it is sufficient to work in Euclidean space. The logarithm of the functional determinant of $\mathcal{M}$ determines the contribution of quarks to the effective action, i.e.\ the generating functional of one-particle irreducible diagrams. We are interested in finite temperature and density, where this determinant can be worked out by standard means of thermal field theory, see, e.g., Ref.\ \cite{Laine:2016hma}. It reads in the Matsubara formalism
\begin{align}
\begin{split}
\ln\det\mathcal{M} = T \sum_{n\in \mathbb{Z}}\int_\bold{p}\tr \ln
M(\nu_n,p)\,,
\end{split}
\end{align}
with the operator 
\begin{align}
\begin{split}
M(\nu_n,p) &=
i\gamma^0(\nu_n-gA_0+h_\omega\omega_0-i\mu)\\
&\quad + i \gamma^j(p_j -g A_j)+m + h_\sigma \sigma\,.
\end{split}
\end{align}
$\nu_n = \pi T(2n+1)$ are the fermionic Matsubara frequencies. The trace in this expression is over color-, flavor- and spinor-indices.

In a thermal medium the temporal gluon fields can have a nonzero VEV $\bar A_0 = \langle A_0 \rangle$ which is directly related to nontrivial VEV of the Polyakov loop \cite{Fukushima:2017csk}. Owing to the invariance under background gauge transformations, the background field can always be rotated into the Cartan subalgebra of the gauge group where it is diagonal diagonal,
\begin{align}\label{eq:A0}
\bar A_0 = \bar A_0^{(3)} T_3 + \bar A_0^{(8)} T_8\,,
\end{align}
with $T_a = \frac{\lambda_a}{2}$ and the Gell-Mann matrices $\lambda_a$. In contrast, we can assume that the spatial gluon field has no VEV. On such a background, the quark propagator is given by
\begin{align}\label{eq:qprop}
G_\psi(\nu_n,\bold{p}) = \frac{-i\gamma^0(\nu_n-g\bar A_0 -i\bar\mu)-i\gamma^j p_j + m_q}{(\nu_n-g\bar A_0 -i\bar\mu)^2 + \bold{p}^2 + m_q^2}\,,
\end{align}
where this is understood to be a diagonal matrix in color space with the diagonal elements of the gauge field background $\bar A_0$. We defined the constituent quark mass
\begin{align}
m_q = m + h_\sigma \bar\sigma\,,
\end{align}
and the effective chemical potential
\begin{align}
\begin{split}
\bar\mu = \mu + i h_\omega\bar\omega_0\,.
\end{split}
\end{align}
To see how mixing arises, it is instructive to consider a diagrammatic expansion first. To this end, we define the vertices
\begin{align}
\Gamma_{\bar\psi \phi_i \psi} =  \frac{\delta M}{\delta\phi_i(0)}\bigg|_{\rm EoM}\,.
\end{align}
Since we are interested in the soft modes relevant for criticality, we can restrict our analysis to zero momentum exchange. Hence, the functional derivatives are carried out at zero momentum and the delta distributions are omitted in favor of explicitly enforcing momentum conservation for all diagrams. 
The direct coupling between any two fields or operators $\phi_i$ and $\phi_j$ at zero momentum exchange generated from the quark determinant is then given by
\begin{align}\label{eq:lindiag}
\begin{split}
\Gamma_{\phi_i\phi_j} &= \frac{\delta^2 \ln\det\mathcal{M}}{\delta \phi_i(0) \delta \phi_j(0)}\bigg|_{\rm EoM}\\
&= T \sum_{n\in \mathbb{Z}}\int_{\bold{p}}\tr\, \Gamma_{\bar\psi \phi_i \psi} G_\psi(\nu_n,\bold{p}) \Gamma_{\bar\psi \phi_j \psi} G_\psi(\nu_n,\bold{p})\,. 
\end{split}
\end{align}
From this expression we already see that certain linear couplings cannot be generated. This includes a mixing of the spatial gluons and the chiral condensate, $\Gamma_{\sigma A_j} = 0$: With the vertices $\Gamma_{\bar\psi \sigma \psi} = h_\sigma$, $\Gamma_{\bar\psi A_j^a \psi} = -igT^a \gamma^j$ and the propagator in \Eq{eq:qprop} it is evident that either the color or the spinor trace in \Eq{eq:lindiag} vanishes. The same reasoning leads to the absence of mixing between the $\omega_0$ and spatial gluon fields. Also the coupling between the chiral condensate and pions, $\Gamma_{\sigma\pi}$, vanishes since scalars cannot be transformed into pseudoscalars by quark exchange. Lastly, the mixing with spatial isoscalar vector mesons, $\Gamma_{\sigma\omega_i}$, can only occur at finite spatial momentum exchange.

In contrast, fields that couple to quarks similar to a chemical potential with a vertex $\sim \gamma_0$ can mix with the chiral condensate at vanishing spatial momentum. After the spinor trace, the integrand of \Eq{eq:lindiag} then is proportional to $(\nu_n-g\bar A_0 - i\bar\mu)$, which can lead to mixing in a medium.

Since the mixing we are considering here involves composite operators, a saddle point approximation is most convenient to extract the relevant linear couplings. It is sufficient to consider the contributions to the effective action from the quark determinant that arise from fluctuations $\delta\sigma$, $\delta\omega_0$ and $\delta A_0$ around the background fields $\bar\sigma$, $\bar\omega_0$ and $\bar A_0$. To leading order, the quark determinant then is
\begin{align}\label{eq:detM}
\begin{split}
\frac{T}{\mathcal{V}}\ln\det\mathcal{\bar M} &=
-2T N_f\int_\bold{p}\bigg\{ \ln\Big[1+N_c\Phi e^{-(E_\bold{p}(\bar \sigma)-\bar\mu)/T}\\
&\quad+N_c\bar\Phi e^{-2(E_\bold{p}(\bar \sigma)-\bar\mu)/T}+e^{-3(E_\bold{p}(\bar \sigma)-\bar\mu)/T}\Big]\\
&\quad + \ln\Big[1+N_c\bar\Phi e^{-(E_\bold{p}(\bar\sigma)+\bar\mu)/T}\\
&\quad+N_c\Phi e^{-2(E_\bold{p}(\bar\sigma)+\bar\mu)/T}+e^{-3(E_\bold{p}(\bar\sigma)+\bar\mu)/T}\Big]\bigg\}\,,
\end{split}
\end{align}
where $\mathcal{V}$ is the spatial volume. We neglected a vacuum contribution as it is irrelevant for our argument. Instead of the components of $\bar A_0$, it can be advantageous to use the temporal Wilson line,
\begin{align}\label{eq:P}
P(\bold{x}) = \mathcal{P} \exp\Bigg[ig\! \int_0^\beta\!dx_0\, A_0(x_0,\bold{x}) \Bigg]\,,
\end{align}
where $x_0$ is the Euclidean (imaginary) time coordinate and $\mathcal{P}$ denotes path ordering. The Polyakov loops are defined as
\begin{align}\label{eq:PL}
\begin{split}
\Phi = \frac{1}{N_c} \langle\tr\, P\rangle\,,\quad
\bar\Phi = \frac{1}{N_c} \langle\tr\, P^\dagger\rangle\,.
\end{split}
\end{align}
For details see, e.g., Ref.\ \cite{Fukushima:2017csk}.
The color trace on the $A_0$ background in the Cartan-subalgebra leads to the Polyakov loops in \Eq{eq:detM}. Due to charge conjugation symmetry breaking, one has $\Phi \neq\bar\Phi$ at finite chemical potential.

By replacing $\bar\phi \rightarrow \bar\phi + \delta \phi$ with $\phi = (\sigma,\Phi,\bar\Phi,\omega_0)$, taking derivatives with respect to $\delta \phi$ and $\delta \tilde\phi$, where $\tilde\phi = (\sigma,\bar\Phi,\Phi,\omega_0)$, and then setting the fluctuating fields to zero, we can extract all the linear couplings between these operators at zero momentum exchange.

Treating the fluctuations requires some caution here. For the saddle point approximation to be well-defined, fluctuations must be in the direction of the steepest descent of the effective action. Put differently, the fluctuations $\delta\phi$ have to be along the stable Lefshetz thimble connected to $\bar\phi$ in field space, see, e.g., \cite{Tanizaki:2015gpl, Alexandru:2020wrj}. The chiral condensate determines the constituent quark masses, so it is real. The scalar field fluctuations around it must be real as well. Similarly, a real free energy requires the Polyakov loops to be real for real chemical potentials. Note that the underlying eigenvalues of the gluon background field need to be complex in this case, see \cite{Dumitru:2005ng, Nishimura:2014rxa, Tanizaki:2015pua, Pisarski:2016ixt} for related discussions. Regarding the vector meson, we have argued that the background field $\bar\omega_0$ needs to be pure imaginary in order for the density to be real. Since $\omega_0$ acts like a chemical potential, the effective action stays real for fluctuations $\delta\omega_0$ in either imaginary or real direction. Inspection of the quark determinant in \Eq{eq:detM} reveals that the dual (or unstable) thimble related to $\bar\omega_0$ is along the imaginary axis, while the stable thimble goes through $\bar\omega_0$ parallel to the real axis, cf.\ also Ref.\ \cite{Mori:2017zyl}. Thus, we need to consider fluctuations of $\omega_0$ in real direction around an imaginary background.

Proceeding with the saddle point approximation, we find for the mixing between the chiral condensate and $\omega_0$,
\begin{align}\label{eq:Gsiom}
    \begin{split}
\Gamma_{\sigma\omega_0} 
&=
-4 N_f T \sum_{n\in \mathbb{Z}}\int_{\bold{p}}\tr_c \frac{(\nu_n-g\bar A_0 -i\bar\mu)\, i h_\omega h_\sigma m_q}{\big[(\nu_n-g\bar A_0 -i\bar\mu)^2 + \bold{p}^2 + m_q^2\big]^2}\\
&=
-4 i N_f N_c h_\omega h_\sigma m_q \int_{\bold{p}} \frac{\partial}{\partial E_\bold{p}^2}\Big[
N_F(E_\bold{p})- \bar N_F(E_\bold{p})
\Big]\,,
    \end{split}
\end{align}
where $\tr_c$ denotes the color trace. 
We have furthermore introduced the modified distribution functions
\begin{align}
    \begin{split}
        N_F(E) &= \frac{1}{N_c} \tr_c n_F(E-\bar\mu+i\bar A_0)\\
        &= \frac{1+2\bar\Phi e^{(E-\bar\mu)/T}+\Phi e^{2(E-\bar\mu)/T}}{1+N_c\bar\Phi e^{(E-\bar\mu)/T}+N_c\Phi e^{2(E-\bar\mu)/T}+e^{3(E-\bar\mu)/T}}\,,\\
        \bar N_F(E) &= \frac{1}{N_c} \tr_c n_F(E+\bar\mu-i\bar A_0)\\
        &= \frac{1+2\Phi e^{(E+\bar\mu)/T}+\bar\Phi e^{2(E+\bar\mu)/T}}{1+N_c\Phi e^{(E+\bar\mu)/T}+N_c\bar\Phi e^{2(E+\bar\mu)/T}+e^{3(E+\bar\mu)/T}}\,,
    \end{split}
\end{align}
where $n_F(x) = (e^{x/T}+1)^{-1}$ is the Fermi-Dirac distribution, and we have defined the quark energy $E_\bold{p} =  \sqrt{\bold{p}^2+m_q^2}$.
At $\bar\mu = 0$, Polyakov loop and anti-loop are identical, $\Phi = \bar\Phi$, and $\Gamma_{\sigma\omega_0}$ vanishes for all temperatures. At $T=0$, this can already be inferred from the first line in \Eq{eq:Gsiom} which then becomes  an odd function of frequency for $\mu=0$. At finite density the degeneracy between $\Phi$ and $\bar\Phi$ is lifted and we conclude that $\Gamma_{\sigma\omega_0}\neq 0$, if $\mu\neq0$.
As a consequence of the repulsive nature of the vector interaction, this coupling is imaginary. We will comment on the physical consequences of this in the next section. That such a mixing arises from the quark determinant in the presence of scalar and vector interaction channels has already been noted in Ref.\ \cite{Kunihiro:1991qu} in the context of the NJL model. Its presence in nuclear matter has been recognized even earlier, e.g., in Ref.\ \cite{Lim:1989ma}.

Next, we investigate the linear coupling between the Polyakov loops and the chiral condensate, $\Gamma_{\sigma \Phi}$ and $\Gamma_{\sigma \bar\Phi}$. 
This is related to fluctuations of $A_0$. Because the quark propagator in \Eq{eq:qprop} is diagonal in color, only temporal gluons with nonzero diagonal elements can contribute.  Without loss of generality, we can therefore assume that $A_0$ stays in the Cartan-subalgebra, and consider fluctuations in the Polyakov loops for simplicity instead. These lead to
\begin{align}\label{eq:sigmaPhi}
\begin{split}
\Gamma_{\sigma\Phi} &= 2 N_f N_c h_\sigma m_q \int_\bold{p}\frac{1}{E_\bold{p}} \bigg\{
e^{2(E_\bold{p}-\bar\mu)/T}\, K(\Phi,\bar\Phi)\\
&\quad\times \Big[N_F(E_\bold{p})-N_c N_F(E_\bold{p})^2 \Big]\\
&\quad+e^{(E_\bold{p}+\bar\mu)/T}\, K\big(\bar\Phi,\Phi\big)\,\Big[2\bar N_F(E_\bold{p})-N_c \bar N_F(E_\bold{p})^2 \Big]
\bigg\}\,,
\end{split}
\end{align}
where we have defined
\begin{align}
K(\Phi,\bar\Phi) = \frac{1}{1+2\bar\Phi e^{(E_\bold{p}-\bar\mu)/T}+ \Phi e^{2(E_\bold{p}-\bar\mu)/T}}\,.
\end{align}
The corresponding coupling between the anti-loop and the chiral condensate is analogously given by
\begin{align}\label{eq:sigmaPhibar}
\begin{split}
\Gamma_{\sigma\bar\Phi} &=2 N_f N_c h_\sigma m_q \int_\bold{p}\frac{1}{E_\bold{p}} \bigg\{
e^{2(E_\bold{p}-\bar\mu)/T}\, K(\Phi,\bar\Phi)\\
&\quad\times \Big[2N_F(E_\bold{p})-N_c N_F(E_\bold{p})^2 \Big]\\
&\quad+e^{(E_\bold{p}+\bar\mu)/T}\,K(\bar\Phi,\Phi)\, \Big[\bar N_F(E_\bold{p})-N_c \bar N_F\big(E_\bold{p})^2 \Big]
\bigg\}\,.
\end{split}
\end{align}
This shows that chiral condensate and temporal gluon fields mix at finite temperature. The argument is sufficiently general to apply to QCD. 
Note that the two mixing terms in \eqref{eq:sigmaPhi} and \eqref{eq:sigmaPhibar} are different at finite $\mu$.

The linear couplings to the chiral condensate are proportional to the quark mass. Thus, the in-medium mixing of the chiral condensate with other modes vanishes in the chiral limit. This follows directly from chiral symmetry.

We also find a nontrivial mixing between $\omega_0$ and the Polyakov loops, 
\begin{align}\label{eq:omegaPhi}
\begin{split}
\Gamma_{\omega_0\Phi} &=-2 iN_f N_c h_\omega \int_\bold{p}\bigg\{
e^{(E_\bold{p}-\bar\mu)/T}\, K(\Phi,\bar\Phi)\\
&\quad\times \Big[ N_F(E_\bold{p})-N_c N_F(E_\bold{p})^2 \Big]\\
&\quad-e^{2(E_\bold{p}+\bar\mu)/T}\,K(\bar\Phi,\Phi)\, \Big[2\bar N_F(E_\bold{p})-N_c \bar N_F(E_\bold{p})^2 \Big]
\bigg\}\,,
\end{split}
\end{align}
and
\begin{align}\label{eq:omegaPhibar}
\begin{split}
\Gamma_{\omega_0\bar\Phi} &=-2 i N_f N_c h_\omega \int_\bold{p}\bigg\{
e^{2(E_\bold{p}-\bar\mu)/T}\, K(\Phi,\bar\Phi)\\
&\quad\times \Big[2N_F(E_\bold{p})-N_c N_F(E_\bold{p})^2 \Big]\\
&\quad-e^{(E_\bold{p}+\bar\mu)/T}\, K(\bar\Phi,\Phi)\, \Big[\bar N_F(E_\bold{p})-N_c \bar N_F\big(E_\bold{p})^2 \Big]
\bigg\}\,.
\end{split}
\end{align}
Note that these two couplings are pure imaginary and unequal at finite $\mu$ as well. 

Finally, there are "off-diagonal" couplings between the Polyakov loops themselves as well,
\begin{align}\label{eq:GPP}
\begin{split}
\Gamma_{\Phi\Phi}&=2 N_f N_c^2 T \int_\bold{p}\bigg\{ 
e^{4(E_\bold{p}-\bar\mu)/T}\, K(\Phi,\bar\Phi)^2\, N_F(E_\bold{p}\big)^2\\
&\quad+ e^{2(E_\bold{p}+\bar\mu)/T}\, K(\bar\Phi,\Phi)^2\, \bar N_F(E_\bold{p})^2
\bigg\}\,,
\end{split}
\end{align}
and
\begin{align}\label{eq:GPbPb}
\begin{split}
\Gamma_{\bar\Phi\bar\Phi}&=2 N_f N_c^2 T \int_\bold{p}\bigg\{ 
e^{2(E_\bold{p}-\bar\mu)/T}\, K(\Phi,\bar\Phi)^2\, N_F(E_\bold{p})^2\\
&\quad+ e^{4(E_\bold{p}+\bar\mu)/T}\, K(\bar\Phi,\Phi)^2\, \bar N_F(E_\bold{p})^2
\bigg\}\,.
\end{split}
\end{align}
Again, these two couplings are identical at $\mu=0$, but this degeneracy is lifted at nonzero $\mu$. 
This exhausts all mixing terms considered here.

The diagonal elements of the two-point function, $\Gamma_{\sigma\sigma}$, $\Gamma_{\omega_0\omega_0}$ and $\Gamma_{\Phi\bar\Phi}$, are also nontrivial and we list them here for completeness:
\begin{align}
\nonumber
\Gamma_{\sigma\sigma} &= 2 N_f N_c h_\sigma^2 \int_{\bold{p}}\Bigg\{ \Bigg(\frac{1}{E_\bold{p}}-\frac{m_q^2}{E_\bold{p}^3}+\frac{m_q^2}{E_\bold{p}^2}\frac{\partial}{\partial E_\bold{p}}\Bigg)\\
\nonumber
&\quad\times\Big[
N_F(E_\bold{p})+ \bar N_F(E_\bold{p})\Big]
\Bigg\}\,,\\[1ex]
\nonumber
\Gamma_{\omega_0\omega_0} &= -2 N_f N_c h_\omega^2 \int_{\bold{p}} \frac{\partial}{\partial E_\bold{p}}\Big[
N_F(E_\bold{p})+ \bar N_F(E_\bold{p})
\Big]\,,\\[1ex]
\nonumber
\Gamma_{\Phi\bar\Phi} &= 2 N_f N_c^2 T \int_\bold{p}\bigg\{ 
e^{3(E_\bold{p}-\bar\mu)/T}\, K(\Phi,\bar\Phi)^2\, N_F(E_\bold{p})^2\\
&\quad+ e^{3(E_\bold{p}+\bar\mu)/T}\, K(\bar\Phi,\Phi)^2\, \bar N_F(E_\bold{p})^2
\bigg\}\,.
\end{align}
In summary, we have shown that QCD in a dense medium features a mixing of the chiral condensate, the quark density and the Polyakov loops. Hence, the physical degrees of freedom have to be mixtures of these (composite) fields. The physical relevance of this observation is discussed in the remainder of this work.

As discussed above, the VEV of $\omega_0$ is directly related to the density of the system, see \Eq{eq:baromega}. Thus, with $\Gamma_{\sigma\omega_0}$ in \Eq{eq:Gsiom} we have provided a microscopic derivation of the mixing between the chiral condensate and the density. The relevance of this mixing for the physics of the QCD critical point has previously been discussed based on phenomenological arguments \cite{Fujii:2004jt, Son:2004iv}. Our analysis shows that this might not be the whole story, however, as further linear couplings to the chiral condensate emerge in QCD.

The mixing with the Polyakov loops is particularly interesting. $\Phi$ and $\bar\Phi$ are related to the (exponential of) the free energy of single, static quarks and antiquarks in the system. Hence, their admixture to the chiral condensate links confinement and chiral symmetry breaking. Furthermore, the temporal gluon field can be viewed as a Lagrange multiplier to enforces Gauss's law in QCD. The mixing with the density therefore leads to the familiar contribution of the quarks to the color charge density. The mixing with $\sigma$ shows that the chiral condensate also contributes to this charge density.

%I am thinking about an object like this as part of the effective action,
%\begin{align}
%   \Gamma \supset \int_x\bar\psi_c(\tau,\bold{x})\, e^{i g \int_\tau^{\tau+\beta}d\tau' A_0^{dd}(\tau',\bold{x})}\, \psi_c(\tau+\beta,\bold{x})\,, 
%\end{align}
%where $c,d$ are color indices, so $A_0^{dd}$ indicates the color trace, and $\psi(\tau+\beta,\bold{x}) = -\psi(\tau,\bold{x})$. This object is a special case of two charges placed at $x$ and $y$, connected by a Wilson line, $\bar\psi_c(x) \,W^{cd}(x,y)\,\psi_d(y)$. Gauss's law is the EoM for $A_0$,
%\begin{align}
%    \frac{\delta\Gamma}{\delta A_0} \sim D_i F^{i0} - \rho = 0\,.
%\end{align}
%$\rho$ is the color charge density. It gets its usual contribution from the conserved color charge current of the quarks, $\rho \supset \psi^\dagger\psi \sim \omega_0$. But now we get another contribution,
%\begin{align}
%    \rho \supset i g \bar\psi_c(t,\bold{x})\, e^{i g \int_t^{t+\beta}d\tau A_0^{dd}(\tau,\bold{x})}\, \psi_c(t+\beta,\bold{x})\,.
%\end{align}

In-medium mixing in QCD is  of course well known. For example the mixing between the vector and axial vector mesons due to interactions with pions in the heat bath \cite{Dey:1990ba} $\rho$-$\omega$ mixing \cite{OConnell:1995nse}, or the mixing between different Goldstone bosons at large baryon-, isospin- and strangeness chemical potentials \cite{Kogut:2001id}. In addition, our discussion is limited to the case of two degenerate quark flavors applied to critical physics. Following the same line of reasoning as above, it is straightforward to see that in general mixing in QCD is considerably more extensive. For example, taking into account the strange quark gives rise to additional mixing with the strange chiral condensate, or, in terms of mesons, a combination of $f_0(500)$ and $f_0(980)$. At finite spatial momentum, also the spatial components of $\omega_\mu$ contribute. If flavor symmetries are relaxed, e.g., through different chemical potentials, and also excited states are taken into account, the mixing proliferates further.

%%%%%%%%%%%%%%%%%%%%%%%%%%%%%%%%%%%%%%%%%%%%%%%%%%
%%%%%%%%%%%%%%%%%%%%%%%%%%%%%%%%%%%%%%%%%%%%%%%%%%
\subsection{Induced Hessian}\label{sec:hess}
%%%%%%%%%%%%%%%%%%%%%%%%%%%%%%%%%%%%%%%%%%%%%%%%%%
%%%%%%%%%%%%%%%%%%%%%%%%%%%%%%%%%%%%%%%%%%%%%%%%%%

We have derived the medium induced mixing in terms of fundamental objects in QCD. But while the chiral condensate and the quark density can be identified with the vacuum expectation values of mesonic operators, the Polyakov loop itself cannot be interpreted as some sort of field. However, since it is the trace of an exponential of an $SU(N_c)$ matrix, cf.\ Eqs.\ \eq{eq:P} and \eq{eq:PL}, we may express it in terms of the eigenvalues $\theta_c$ of this matrix,
\begin{align}
\Phi = \frac{1}{N_c} \sum_{c=1}^{N_c}\Big\langle e^{ig\theta_c/T} \Big\rangle\,.
\end{align}
The eigenvalues obey $\sum_c \theta_c = 0$, so there are $N_c-1$ independent eigenvalues. We use the following parametrization for QCD,
\begin{align}
    \theta_1 = \frac{a_3}{2}+\frac{a_8}{2\sqrt{3}}\,,\quad
    \theta_2 = -\frac{a_3}{2}+\frac{a_8}{2\sqrt{3}}\,,\quad
    \theta_3 = -\frac{a_8}{\sqrt{3}}\,,
\end{align}
which leads to
\begin{align}\label{eq:PLpara}
    \begin{split}
        \Phi &= \frac{1}{3} e^{\frac{i a_8}{2 \sqrt{3} T}}\bigg[2\cos\bigg(\frac{a_3}{2T}\bigg) + e^{-\frac{3 i a_8}{2 \sqrt{3} T}}\bigg]\,,\\
        \bar\Phi &= \frac{1}{3} e^{\frac{-i a_8}{2 \sqrt{3} T}}\bigg[2\cos\bigg(\frac{a_3}{2T}\bigg) + e^{\frac{3 i a_8}{2 \sqrt{3} T}}\bigg]\,.
    \end{split}
\end{align}
This parametrization can be motivated by using that the Polyakov loop is related to a nonvanishing temporal gluon background field $\bar A_0$. This field can be rotated into the Cartan subalgebra, see \Eq{eq:A0}, and we can identify the components of $\bar A_0$ with our parameters, $g\bar A_0^{(3)} = a_3$ and $g\bar A_0^{(8)} = a_8$. Hence, $a_3$ and $a_8$ can loosely be interpeted as eigenvalues of a temporal gluon background field. As has been noted previously, real Polyakov loops require a purely imaginary $a_8$ at finite density \cite{Dumitru:2005ng, Nishimura:2014rxa, Tanizaki:2015pua, Pisarski:2016ixt}. We emphasize that the Polyakov loop is not the same as the exponential of $\bar A_0$, although both are linked to the confinig properties of QCD and can serve as order parameters for confinement in the pure gauge theory \cite{Braun:2007bx}. 

Most importantly, the parametrization \Eq{eq:PLpara} provides a natural way to express the Polyakov loops in terms of field-like variables $a_3$ and $a_8$. We can use this to define the Hessian, or mass matrix, of the QCD effective action,
\begin{align}\label{eq:Hgen}
H = \big(\Gamma_{\phi_i\phi_j}\big)\,,
\end{align}
which is understood to be a $4\!\times\!4$ matrix with $\phi= (\sigma, a_3, a_8, \omega_0)$. The entries involving $a_3$ and $a_8$ are directly related to the couplings involving $\Phi$ and $\bar\Phi$ via
\begin{align}\label{eq:PLconv}
        \Gamma_{\phi_i a_{3/8}} = \bigg(\frac{\partial\Phi}{\partial a_{3/8}}\bigg)\, \Gamma_{\phi_i\Phi} + \bigg(\frac{\partial\bar\Phi}{\partial a_{3/8}}\bigg)\, \Gamma_{\phi_i\bar\Phi}\,.
\end{align}
With this, the Hessian in \Eq{eq:Hgen} is, to leading order in the saddle point approximation, determined by the linear couplings derived in \Sec{sec:mix}.
It follows from the analysis of the previous section that $H$ has nonzero off-diagonal elements in a medium. This implies that diagonalization is required and the physical degrees of freedom are linear combinations of these fields.

The nontrivial mixing found here is directly related to the breaking of charge conjugation symmetry, $\mathcal{C}$, at finite density.
Since $\omega_0$ is a component of a vector field, under charge conjugation it transforms as $\mathcal{C}\omega_0 = - \omega_0$. Because gauge fields transform as $\mathcal{C} A_\mu = - A_\mu^t$, under $\mathcal C$ the parameter $a_8$ changes sign\footnote{Note that the Polyakov loops are even functions of $a_3$.} and $\Phi$ and $\bar\Phi$ are exchanged.
Nonzero $\Gamma_{\sigma\omega_0}$, $\Gamma_{\omega_0\Phi}$ and $\Gamma_{\omega_0\bar\Phi}$ therefore break $\mathcal{C}$-symmetry, reflecting the $\mathcal{C}$-symmetry breaking at finite density. $\mathcal{C}$ symmetry breaking also leads to $\Phi\neq\bar\Phi$. For the mixing terms this implies $\Gamma_{\sigma\Phi} \neq \Gamma_{\sigma\bar\Phi}$, $\Gamma_{\omega_0\Phi} \neq \Gamma_{\omega_0\bar\Phi}$
and $\Gamma_{\Phi\Phi} \neq \Gamma_{\bar\Phi\bar\Phi}$. This also implies that $a_8$ in \Eq{eq:PLpara} must be nonzero and imaginary at finite density, because the Polyakov loops are real. Hence, all off-diagonal terms of the Hessian in \Eq{eq:Hgen} involving either a repulsive $\omega_0$ or $a_8$ are purely imaginary at finite density and change sign under $\mathcal{C}$.

This shows that the Hessian becomes non-Hermitian if charge conjugation symmetry is broken. However, complex conjugation $\mathcal{K}$ exchanges $\Phi$ and $\bar\Phi$ by definition. The Polyakov loops therefore transform into themselves under $\mathcal{CK}$. Since all couplings with either a single $\omega_0$ or a single $a_8$ are imaginary, they change sign under $\mathcal{K}$. Putting  all this together, we see that the system retains an antilinear $\mathcal{C}\mathcal{K}$ symmetry at finite density \cite{Nishimura:2014kla}. We will get back to this point in the next section.

%%%%%%%%%%%%%%%%%%%%%%%%%%%%%%%%%%%%%%%%%%%%%%%%%%
%%%%%%%%%%%%%%%%%%%%%%%%%%%%%%%%%%%%%%%%%%%%%%%%%%
\section{Mixing and the phase structure}\label{sec:phase}
%%%%%%%%%%%%%%%%%%%%%%%%%%%%%%%%%%%%%%%%%%%%%%%%%%
%%%%%%%%%%%%%%%%%%%%%%%%%%%%%%%%%%%%%%%%%%%%%%%%%%

In-medium mixing can have large effects on the phase structure. We show that, firstly, it changes the nature of the soft modes that govern the system near second-order phase transitions. Secondly, the resulting "weaker" (as compared to hermiticity) $\mathcal{CK}$ symmetry of the Hessian can give rise to spatially modulated regimes.

%%%%%%%%%%%%%%%%%%%%%%%%%%%%%%%%%%%%%%%%%%%%%%%%%%
%%%%%%%%%%%%%%%%%%%%%%%%%%%%%%%%%%%%%%%%%%%%%%%%%%
\subsection{Criticality}\label{sec:crit}
%%%%%%%%%%%%%%%%%%%%%%%%%%%%%%%%%%%%%%%%%%%%%%%%%%
%%%%%%%%%%%%%%%%%%%%%%%%%%%%%%%%%%%%%%%%%%%%%%%%%%

If the order parameter is known, like the chiral condensate $\bar\sigma$ for the chiral phase transition, the second-order critical point can be identified, e.g., through the divergence of the slope of $\bar\sigma(T)$. While this is sufficient to locate the critical point, it says nothing about its nature. In order to identify the critical modes, we use that, following the work of Lee and Yang, a branch point of the thermodynamic potential $\bar\Omega(\mu)$, where we omit other arguments such as $T$ since they are irrelevant here, is expected to pinch the real chemical potential axis at the CEP \cite{Yang:1952be, Lee:1952ig}. This branch point is called the Yang-Lee edge singularity (YLE). Away from the phase transition, e.g., at small $\mu$ in QCD, the YLE is the singularity in the complex $\mu$ plane closest to the real $\mu$ axis. In QCD, we identify the YLE as follows:
We assume that the effective potential $\Omega(\phi)$ is continuously differentiable and the EoM,
\begin{align}\label{eq:EoM}
\frac{\partial \Omega(\phi)}{\partial \phi_i}\bigg|_{\phi_i=\bar\phi_i}=0\,,
\end{align}
has solutions for all positive $T$ and complex $\mu$.
The effective potential evaluated on the solution of the EoM defines the thermodynamic potential,
\begin{align}
\bar\Omega(\mu) = \Omega\big(\bar\phi(\mu)\big)\,.
\end{align}
The implicit function theorem states that \Eq{eq:EoM} can locally be inverted to give unique continuously differentiable functions $\bar\phi_i(\mu)$ as long as the Hessian
\begin{align}\label{eq:hess}
H(\mu) = \Bigg(\frac{\partial^2 \Omega}{\partial \phi_i \partial \phi_j} \Bigg)\bigg|_{\phi_i=\bar\phi_i}
\end{align}
is invertible. Conversely, if $H(\mu)$ has at least one vanishing eigenvalue at $\mu_{\rm YLE} \in \mathbb{C}$, then $\bar\phi_i(\mu)$ is not differentiable at $\mu_{\rm YLE}$. In presence of a branch cut, this singularity corresponds to the branch point. Thus, the YLE is identified as the (complex) chemical potential, $\mu_{\rm YLE}$, where at least one of the eigenvalues of the Hessian becomes zero.

The mixing discussed in the previous section gives rise to a nontrivial Hessian which, for two degenerate quark flavors at vanishing momentum, is a $4\times 4$ matrix. Since we have shown that the off-diagonal entries are in general nonzero in the medium, it follows that neither the chiral condensate $\sigma \sim \langle \bar\psi \psi \rangle$ alone, nor a linear combination between only the condensate and the baryon density $\sim \omega_0 $ can be the true critical mode of the CEP in QCD. It should rather be a mixture of the  $\sigma $ field, the conserved isoscalar density $\omega_0$, and the Polyakov loops here representing the nontrivial electric gluon correlations at finite temperature in QCD. 

Since the couplings $\Gamma_{\sigma\Phi}$ and $\Gamma_{\sigma\bar\Phi}$, or, equivalently, $\Gamma_{\sigma a_3}$ and $\Gamma_{\sigma a_8}$, are nonzero also at $\mu=0$ as long as $T>0$ and $m_q>0$, cf.\ Eqs. \eq{eq:sigmaPhi} and \eq{eq:sigmaPhibar}, also the potential critical mode of the chiral transition at small quark mass has an admixture of the Polyakov loop. This is, of course, only relevant if there is a second order transition at nonzero quark mass\footnote{The nature of the chiral phase transition towards the chiral limit is not entirely settled yet, e.g., see \cite{Kuramashi:2020meg, Cuteri:2021ikv, Dini:2021hug,Fejos:2022mso} for recent discussions.}. If the second order transition happens to be at zero current quark mass, the mixing vanishes.

The response of the system to changes in external parameters, such as $T$, $\mu$ or the current quark mass $m$, is characterized by the susceptibilities. They inherit the singular behavior at the CEP from the YLE. To see this, we note that if $\Omega(\phi)$ is the effective potential and we denote $X=(T,\mu,m)$, the EoM leads to
\begin{align}
\frac{\partial\bar\phi_i}{\partial X_a} = - H_{ij}^{-1}\frac{\partial^2 \Omega}{\partial \phi_j \partial X_a}\bigg|_{\rm EoM}\,.
\end{align}
Using this relation, we can derive for the static susceptibilities,
\begin{align}
\chi_{ab} = \frac{d^2\bar\Omega}{dX_a dX_b}\,,
\end{align}
the identity
\begin{align}\label{eq:susc}
\chi_{ab} = \frac{\partial^2 \bar\Omega}{\partial X_a \partial X_b} + \frac{\partial^2\Omega}{\partial X_a\partial\phi_i} H_{ij}^{-1} \frac{\partial^2\Omega}{\partial\phi_j \partial X_b}\bigg|_{\rm EoM}\,,
\end{align}
see also, e.g., Refs.\ \cite{Fu:2015naa,Pisarski:2016ixt, Almasi:2017bhq}.
The susceptibilities diverge at the CEP as a result of the YLE, because $\det\, H = 0$ at the YLE and $H^{-1} \sim (\det\, H)^{-1}$.
Since the susceptibilities are only sensitive to $\det\, H$, not the individual eigenvalues, they in general cannot reveal the nature of the critical mode.

However, in the vicinity of a second-order transition universality entails that the system can be described solely in terms of the soft modes of the system. The critical modes are a crucial, and in the static case the only, part of these modes. As mentioned above, their nature determines the universal properties of the system.

%%%%%%%%%%%%%%%%%%%%%%%%%%%%%%%%%%%%%%%%%%%%%%%%%%
%%%%%%%%%%%%%%%%%%%%%%%%%%%%%%%%%%%%%%%%%%%%%%%%%%
\subsection{Modulations and instabilities}\label{sec:mod}
%%%%%%%%%%%%%%%%%%%%%%%%%%%%%%%%%%%%%%%%%%%%%%%%%%
%%%%%%%%%%%%%%%%%%%%%%%%%%%%%%%%%%%%%%%%%%%%%%%%%%

In addition to the physics near critical points, we have already seen that mixing leads to a non-Hermitian Hessian. 
This has important consequences for the eigenvalues. As noted above, in this case an antilinear $\mathcal{C}\mathcal{K}$ symmetry remains. The Hessian and its complex conjugate then obey the relation
\begin{align}
    H = C H^* C\,,
\end{align}
where $C$ is a diagonal matrix with entries $-1$ for the components corresponding to the fields that lead to imaginary mixing and $+1$ everywhere else. In the notation of \Eq{eq:Hgen}, we have
\begin{align}
    C = {\rm diag}(1,-1,1,-1)\,.
\end{align}
This relation implies that $H$ and $H^*$ have the same set of eigenvalues. The eigenvalues of the Hessian are therefore either real or come in complex conjugate pairs. This can be induced by the competition between repulsive and attractive bosonic interactions or the lifted degeneracy between the Polyakov loop $\Phi$ and its conjugate $\bar\Phi$ at nonzero real baryon chemical potential \cite{Nishimura:2014kla,Nishimura:2014rxa,Nishimura:2015lit,Nishimura:2016yue,Schindler:2019ugo}. The former can be deduced from the imaginary mixing $\Gamma_{\sigma \omega_0}\neq 0$ in \Eq{eq:Gsiom}, and the latter from $\Gamma_{\sigma\Phi} \neq \Gamma_{\sigma\bar\Phi}$ and $\Gamma_{\Phi\Phi} \neq \Gamma_{\bar\Phi\bar\Phi}$, cf.\ Eqs.\ \eqref{eq:sigmaPhi}, \eqref{eq:sigmaPhibar}
and \eq{eq:GPP}, \eq{eq:GPbPb}. 
A combination of these effects emerge in the mixing $\Gamma_{\omega_0\Phi}\neq \Gamma_{\omega_0\bar\Phi}\neq 0$ in Eqs.\ \eq{eq:omegaPhi} and \eq{eq:omegaPhibar}.
All this arises from the breaking of charge conjugation symmetry at $\mu >0$. 

The Hessian discussed here corresponds to the static self-energy corrections at zero momentum. Hence, taking only these corrections into account, the scalar part of the inverse static propagator matrix is
\begin{align}\label{eq:Ginv}
    \big\langle \phi(0,\bold{p}) \phi(0,\bold{0})\big\rangle^{-1} \sim G_\phi^{-1}(\bold{p}) = \bold{p}^2 + H\,,
\end{align}
where $H$ the $n \!\times\! n$ Hessian matrix of the multi-field $\phi$. The static propagator of the eigenmodes $\chi_i$ to the eigenvalue of the Hessian $H_i$ is
\begin{align}
 G_\chi^{-1}(\bold{p}) = \bold{p}^2 + {\rm diag}\,(H_1,\dots, H_n)\,.
\end{align}
Since we have shown that the eigenvalues $H_i$ can in general be complex, the propagator of an eigenmode at large spatial separation $r$ is
\begin{align}\label{eq:Gasymp}
G_{\chi_i}(r)\; \xrightarrow{r\rightarrow\infty}\; \sim e^{- {\rm Re}[H_i]\, r} \sin\big( {\rm Im}[H_i]\, r \big)\,.
\end{align}
Hence, complex eigenvalues lead to spatial modulations in particle correlations. The imaginary part determines the wave number of this modulation. Following our discussion, we expect this to occur at finite density, leading to so-called complex or patterned regimes in the phase diagram \cite{Schindler:2019ugo, Schindler:2021otf}. They are separated from regimes with real eigenvalues by disorder lines \cite{PhysRevB.1.4405,Akerlund:2016myr}.

In addition to the identification of spatial modulations, the eigenvalues of the Hessian can also be used to test the stability of the ground state, which we assumed to be homogeneous here. To see this, note that one-loop corrections to the free energy from fluctuations described by $G_\phi$ in \Eq{eq:Ginv} are given by $\sim \ln\det(\bold{p}^2 + H)$. If the determinant becomes negative, the free energy becomes complex and the system is unstable \cite{Weinberg:1987vp}. The stability of the ground state can hence be tested by studying the poles of the propagator in $\bold{p}^2$. They can be read off from the zeros of the equation 
\begin{align}\label{eq:mesdet}
    \det(\bold{p}^2 + H) = 0\,.
\end{align}
This is a characteristic equation of $H$ with solutions $\bold{p}^2_i = -H_i$, where $i=1,\dots,n$ and $H_i$ are the eigenvalues of $H$. 
We can therefore directly use the eigenvalues of the Hessian to perform a stability analysis. Since this does not take into account momentum-dependent self-energy corrections, such an analysis cannot detect all possible instabilities, cf., e.g., \cite{Buballa:2014tba,  Tripolt:2017zgc, Koenigstein:2021llr, Pannullo:2023one, Pannullo:2023cat, Motta:2023pks}. However, owing to the nontrivial structure of the Hessian in the presence of mixing, certain instabilities can readily be detected. This idea of doing a stability analysis based on mixing has been put forward in Ref.\ \cite{Schindler:2019ugo, Schindler:2021otf}.

If the eigenvalues are either real and positive or, as discussed above, complex, the system remains stable. The former case corresponds to ordinary screening and the latter to spatially modulated correlations in a system with a homogeneous ground state\footnote{A stable homogeneous phase with spatial modulations has also been discussion in Refs.\ \cite{Pisarski:2020dnx, Pisarski:2021aoz}, where it has been dubbed a quantum-pion liquid.}. In both cases, the free energy is real for any real spatial momentum.

Instabilities occur if at least one eigenvalue is negative. Then the static Euclidean propagator has a pole at $\bold{p}_i^2>0$ and the determinant in \Eq{eq:mesdet} is negative for some $\bold{p}^2 \leq 0$. This typically entails that an inhomogeneous ground state with a wave number related to the pole position is favored. Following Ref.\ \cite{Schindler:2021otf}, these instabilities can be classified further: If there is an even number of negative eigenvalues, the propagator is positive at $\bold{p}^2 = 0$ but negative at nonzero momentum. Thus, the system is stable against homogeneous fluctuations, but has an inhomogeneous instability. For an odd number of negative eigenvalues, the propagator is already negative at $\bold{p}^2 = 0$ and the system is unstable against both homogeneous and inhomogeneous fluctuations. These two cases have been called patterned and unstable in Ref.~\cite{Schindler:2021otf}, respectively.

%%%%%%%%%%%%%%%%%%%%%%%%%%%%%%%%%%%%%%%%%%%%%%%%%%
%%%%%%%%%%%%%%%%%%%%%%%%%%%%%%%%%%%%%%%%%%%%%%%%%%
\section{Model Studies}\label{sec:ex}
%%%%%%%%%%%%%%%%%%%%%%%%%%%%%%%%%%%%%%%%%%%%%%%%%%
%%%%%%%%%%%%%%%%%%%%%%%%%%%%%%%%%%%%%%%%%%%%%%%%%%

In order to understand the mixing in QCD and illustrate its physical consequences in more detail, it is instructive to do explicit calculations. To this end, we use well-established low-energy models based on the coupling between quarks, mesons and gluon background fields. We first employ a standard Polyakov--Quark-Meson (PQM) model as the simplest example of the main effects in \Sec{sec:PQM}, and then consider a more complete model, which in addition features the repulsive vector interaction in \Sec{sec:PQMV}.

PQM models are widely used as they describe various features of QCD at low energies, especially regarding thermodynamics and the phase structure, see \cite{Fukushima:2017csk} and references therein.

%%%%%%%%%%%%%%%%%%%%%%%%%%%%%%%%%%%%%%%%%%%%%%%%%%
%%%%%%%%%%%%%%%%%%%%%%%%%%%%%%%%%%%%%%%%%%%%%%%%%%
\subsection{Polyakov--Quark-Meson model}\label{sec:PQM}
%%%%%%%%%%%%%%%%%%%%%%%%%%%%%%%%%%%%%%%%%%%%%%%%%%
%%%%%%%%%%%%%%%%%%%%%%%%%%%%%%%%%%%%%%%%%%%%%%%%%%

Since the admixture of the Polyakov loops in the critical mode is perhaps the most surprising finding of this work, we first study a two-flavor PQM model which describes the low-energy dynamics of quarks, $\sigma$, pions and the Polyakov loops. 
The Euclidean action of this simple PQM model is
\begin{align}\label{eq:SPQM}
\begin{split}
S_{\rm PQM} &=\int_0^\beta\! dx_0 \int\! d^3x\, \Bigg\{ \bar\psi \big[ \gamma^\mu \partial_\mu + \gamma^0 (\mu + i A_0) \big] \psi \\
&\quad + \frac{h_\sigma}{2} \bar\psi \big( \sigma + i \gamma_5 \boldsymbol{\pi}\cdot\boldsymbol{\tau} \big) \psi + \frac{1}{2} (\partial_\mu \sigma)^2 + \frac{1}{2} (\partial_\mu \boldsymbol{\pi})^2\\
&\quad +V(\sigma,\boldsymbol{\pi}) + U(\Phi,\bar\Phi)\Bigg\}\,,
\end{split}
\end{align}
where $x_0$ is imaginary time, $\beta=1/T$ and $\boldsymbol{\tau}$ are the Pauli matrices. Quarks and mesons are coupled with the Yukawa coupling $h$. The meson effective potential,
\begin{align}\label{eq:VS}
V(\sigma,\boldsymbol{\pi}) = \frac{\lambda}{4}\big( \sigma^2 + \boldsymbol{\pi}^2-\nu^2 \big)^2 - j\sigma\,,
\end{align}
consists of a chiral $O(4)$-symmetric part $\sim \lambda$ and a current $j$ that explicitly breaks this symmetry down to $O(3)$. The latter reflects the finite current quark mass and, consequently, a nonzero pion mass in the phase with spontaneously broken chiral symmetry, $j = f_\pi m_\pi^2$, with pion mass and decay constant $m_\pi$ and  $f_\pi$. 

For the Polyakov-loop potential $U(\Phi,\bar\Phi)$, we use a simple $Z(3)$ center-symmetric polynomial,
\begin{align}\label{eq:VP}
\frac{1}{T^4}U(\Phi,\bar\Phi) = -\frac{b_2(T)}{2} \Phi\bar\Phi-\frac{b_3}{3} \big(\Phi^3+\bar\Phi^3\big) + \frac{b_4}{4} (\Phi\bar\Phi)^2\,,
\end{align}
with the following Ansatz for the temperature dependent coefficient
\begin{align}
b_2(T)=  \sum_{n=0}^3 a_n\, \bigg(\frac{T_0}{T}\bigg)^n\,.
\end{align}
This parametrization of the potential has been established in \cite{Ratti:2005jh}, where the parameters were fixed to reproduce the thermodynamics of the pure gauge theory. In particular, $T_0  = 270$\,MeV is the corresponding critical temperature of the first-order deconfinement phase transition in $SU(3)$ pure Yang-Mills theory.

The free energy functional of the PQM model is
\begin{align}
\begin{split}
F &= \ln\int\!\mathcal{D}\phi\, \exp\Big\{-S_{\rm PQM}[\phi]\Big\}\\
 &= \ln\int\!\mathcal{D}(A_0,\sigma,\boldsymbol{\pi})\, \exp\Bigg\{-\int_x \bigg[\frac{1}{2}(\partial_\mu\sigma)^2+\frac{1}{2}(\partial_\mu\boldsymbol{\pi})^2\\
 &\quad + V(\sigma,\boldsymbol{\pi}) + U(\Phi,\bar\Phi)\bigg]
 +\ln\det\, \mathcal{M}\big(A_0,\sigma,\boldsymbol{\pi}\big)\Bigg\}\,,
\end{split}
\end{align}
with $\phi = (\psi,\bar\psi,A_0,\sigma,\boldsymbol{\pi})$. We have integrated out the quarks in the second line, which gives rise to the functional determinant of the Dirac operator,
\begin{align}
\begin{split}
\mathcal{M}\big(A_0,\sigma,\boldsymbol{\pi}\big) &= \gamma^\mu\partial_\mu +\gamma^0\big(\mu + i A_0\big)+\frac{h_\sigma}{2}\big( \sigma + i \gamma_5 \boldsymbol{\pi}\cdot\boldsymbol{\tau} \big)\,.
\end{split}
\end{align}
In mean-field approximation, with the gluon background in the Cartan-subalgebra and assuming a homogeneous VEV $\bar\phi = (0,0,\bar A_0,\bar\sigma,0)$, the effective potential $\Omega = - T F /\mathcal{V}$ becomes
\begin{align}\label{eq:omega}
\begin{split}
\Omega(\sigma,\Phi,\bar\Phi) &= V(\sigma) + U(\Phi,\bar\Phi) - \frac{T}{\mathcal{V}} \ln\det\, \mathcal{M}\big(A_0,\sigma\big)\,.
\end{split}
\end{align}
The functional determinant is identical to the one in \Eq{eq:detM} with the replacement $\bar\mu \rightarrow \mu$, since our model does not have vector mesons.

%%
%%%%%%%%%%
\begin{figure*}
\centering
\includegraphics[width=.46\textwidth]{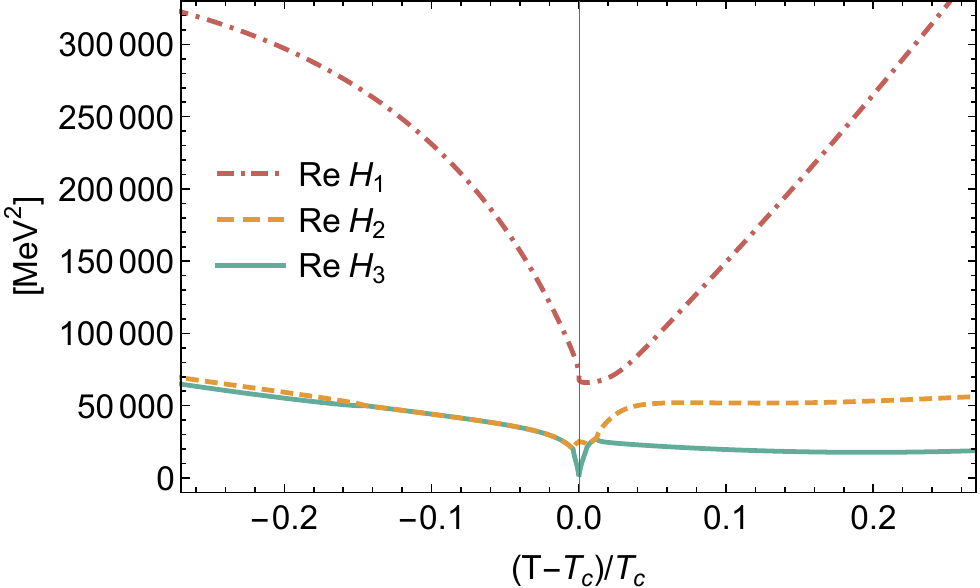}
\hfill
\includegraphics[width=.46\textwidth]{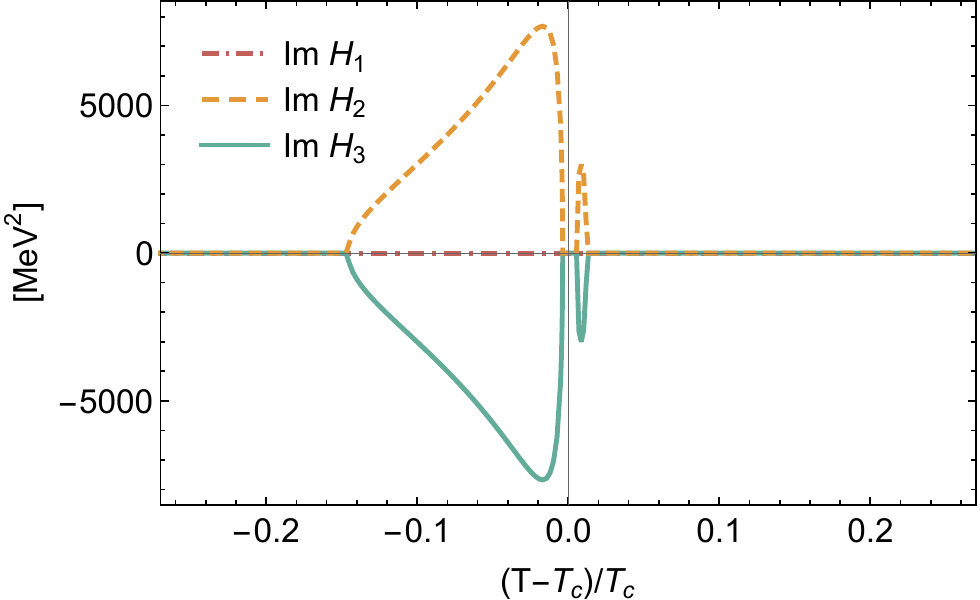}
\caption{Eigenvalues $H_1$, $H_2$ and $H_3$ of the Hessian of the PQM model at the critical chemical potential as functions of $T$. 
The real and imaginary parts of these eigenvalues are shown in the left and right plot respectively.}
\label{fig:EV}
\end{figure*}
%%%%%%%%%%
%%

%%
%%%%%%%%%%
\begin{figure}[b]
\centering
\includegraphics[width=.9\columnwidth]{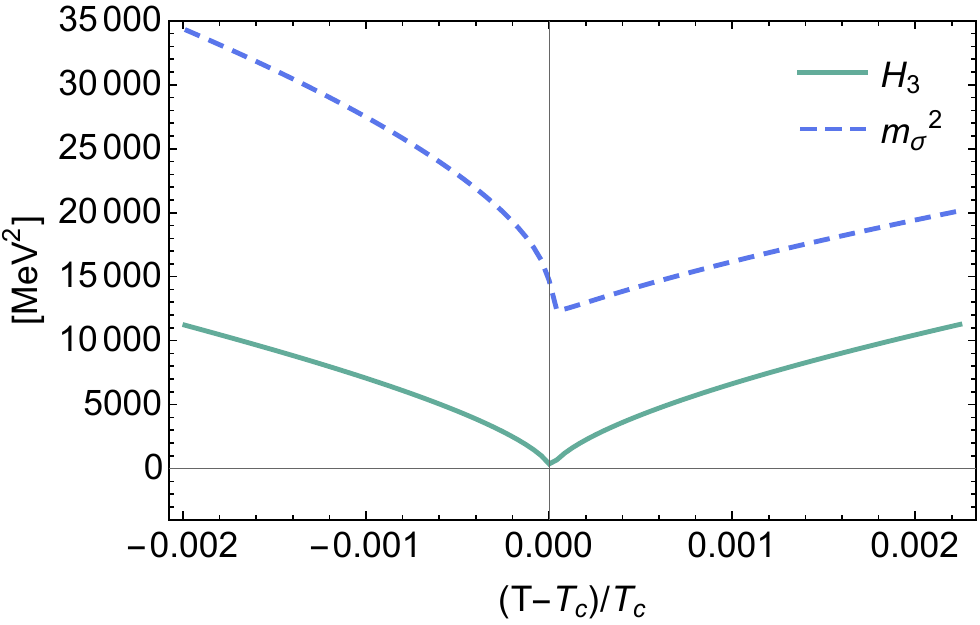}
\caption{Scalar meson mass (dashed) and the eigenvalue $H_3$ of the Hessian (solid) as functions of $T$ close to the CEP. While the scalar meson mass stays finite at the CEP, the eigenvalue vanishes.}
\label{fig:massvsev}
\end{figure}
%%%%%%%%%%
%%

We fix the free parameters of the model as follows. The pion decay constant $f_\pi = 93$\,MeV is identified with $\bar\sigma$. We choose the explicit symmetry breaking such that in vacuum $m_\pi = 138$\,MeV. This fixes the value of the source $j$. The constituent quark mass in vacuum is set to $m_q = \frac{1}{2} h_\sigma f_\pi = 300$\,MeV, which fixes $h_\sigma$. The remaining model parameters are related to the physical quantities as follows: $\nu^2 = f_\pi^2-m_\pi^2/\lambda$ and $\lambda = (m_\sigma^2-m_\pi^2)/ (2 f_\pi^2)$. We choose $m_\sigma = 600$\,MeV, so the remaining parameters are fixed. We emphasize that the latter two relations are strictly mean-field and only valid because we neglect the vacuum contribution of the quark determinant. Using dimensional regularization, this contribution renormalizes the quartic meson coupling in vacuum with a term $\sim \phi^4 \ln (\phi)$. However, since this term has no influence on the mixing, it is irrelevant for our qualitative discussion here. It will be taken into account for completeness in the next section.

The mixing between $\sigma$ and the Polyakov loops is identical to the one in Eqs.\ \eq{eq:sigmaPhi} and \eq{eq:sigmaPhibar}, as it is solely induced by the quark determinant in the PQM. The PQM has the Hessian
\begin{align}\label{eq:HPQM}
H_{\rm PQM} = \begin{pmatrix} \Omega_{\sigma\sigma} & \Omega_{\sigma a_3} & \Omega_{\sigma a_8} \\
\Omega_{\sigma a_3} & \Omega_{a_3  a_3} & \Omega_{a_3 a_8} \\
\Omega_{\sigma a_8} & \Omega_{a_3 a_8} & \Omega_{a_8 a_8}  \end{pmatrix}\,,
\end{align}
with the entries $\Omega_{\phi_i\phi_j} = \frac{\partial^2\Omega}{\partial\phi_i\partial\phi_j}\big|_{\rm EoM} = \Gamma_{\phi_i\phi_j}$ given in \Sec{sec:mix}. The conversion from $\Phi$, $\bar\Phi$ to $a_3$, $a_8$ is done using Eqs.\ \eq{eq:PLpara} and \eq{eq:PLconv}. Since $\Omega_{\sigma a_8}$ and $\Omega_{a_3 a_8}$ are imaginary at $\mu\neq 0$, $H_{\rm PQM}$ is not Hermitian, but $\mathcal{C}\mathcal{K}$ symmetric. Thus, complex eigenvalues are allowed as long as they come in conjugate pairs. The existence of a disorder line is therefore possible in this model, cf.\ \cite{Nishimura:2014rxa} and the discussion in \Sec{sec:mod}.

%%%%%%%%%%%%%%%%%%%%%%%%%%%%%%%%%%%%%%%%%%%%%%%%%%
%\subsubsection{Screening masses and the critical mode}\label{sec:mass}
%%%%%%%%%%%%%%%%%%%%%%%%%%%%%%%%%%%%%%%%%%%%%%%%%%

We focus on the static properties of the model. In order to identify the critical mode, it is important to clarify what the relevant mass is. For a general Euclidean propagator $G_\chi(p_0,\bold{p})$ of a scalar field $\chi$, we can define two masses:
\begin{itemize}
    \item\emph{pole mass}: $G_\chi^{-1}(i m_{\rm pole},\bold{0}) = 0$
    \item\emph{screening mass}: $G_\chi^{-1}(0,i m_{\rm scr} \, \bold{\hat p}) = 0$
\end{itemize}
The former is the mass of the particle that is relevant, e.g., for scattering processes. It represents the pole in the propagator of a stable particle at timelike momenta. The latter formally corresponds to a pole in the propagator at the spatial momentum variable $\bold{p}^2 = - m_{\rm scr}^2$ and determines the large-distance behavior of the correlations in spatial directions. In fact \cite{Zinn-Justin:2002ecy}, the screening mass 
\begin{align}
m_{\rm scr} = {1}/{\xi} 
\end{align}
is the inverse correlation length $\xi$ defined via
\begin{align}\label{eq:mscr}
\lim_{|\bold{x}_1 - \bold{x}_2|\rightarrow \infty}\big\langle \chi(t,\bold{x}_1)\chi(t,\bold{x}_2)\big\rangle \sim e^{- r/\xi}\,,
\end{align}
with $r = |\bold{x}_1 - \bold{x}_2|$.
If $\chi$ is the critical mode, then its correlation length diverges at the CEP. Conversely, its screening mass vanishes. In this case, the order of the limits of temporal and spatial momenta going to zero is crucial. From the definition of the screening mass follows that at the CEP, if $\chi$ is the critical mode,
\begin{align}
\lim_{|\bold{p}|\rightarrow 0}\, \lim_{p_0\rightarrow 0} G_\chi^{-1}(p_0,\bold{p}) = 0\,.
\end{align}
Due to Landau damping in a thermal medium, these limits in general do not commute \cite{Bellac:2011kqa}. In order to extract the screening mass of the critical mode, the $p_0\rightarrow 0$ limit has to be taken first. This entails that we can extract the screening mass in Euclidean space, as it occurs for spacelike momenta. We can express the inverse propagator at $p_0=0$ as the sum of the tree-level contribution and self-energy corrections $\Pi(p_0,\bold{p}^2)$,
\begin{align}
G_\chi^{-1}(0,\bold{p}) = \bold{p}^2 + m_\chi^2 + \Pi(0,\bold{p}^2)\,,    
\end{align}
where we assumed symmetry of the system under spatial rotations. For the critical mode it is therefore sufficient to consider the renormalized mass
\begin{align}
\bar m_\chi^2 = m_\chi^2 + \lim_{|\bold{p}|\rightarrow 0}\Pi(0,\bold{p}^2)\,,
\end{align}
as it coincides with the screening mass at the CEP. This mass is also known as the \emph{curavture mass}. This means that we can directly study the Hessian $H$ in \Eq{eq:Hgen}. From the comparison between Eqs.\ \eq{eq:Gasymp} and \eq{eq:mscr} follows that the curvature mass of the critical mode is given by the eigenvalue of $H$ whose real part vanishes at the CEP. Moreover, from our discussion in \Sec{sec:crit} follows that the imaginary part of the eigenvalue has to vanish as well, as otherwise there would not be an edge singularity. We note that the vicinity of the CEP $\bar m_\chi$ is expected to be very close to the screening mass.

To proceed, we solve the EoM for the effective potential in \Eq{eq:omega}. By evaluating the second derivatives of the effective potential on the solutions $\bar\phi(T,\mu)$ we obtain the Hessian in \Eq{eq:HPQM}.
We identify the CEP through the YLE, i.e. through a vanishing eigenvalue of the Hessian at real chemical potential. With the parameters discussed above, we find
\begin{align}
(T_{\rm CEP}, \mu_{B,{\rm CEP}})\big|_{\rm PQM} \approx (189,493.5)\, {\rm MeV}\,,
\end{align}
where $\mu_B = 3 \mu$ is the baryon chemical potential.

In \Fig{fig:EV} we show the three eigenvalues $H_1$, $H_2$ and $H_3$ at $\mu_B = \mu_{B,{\rm CEP}}$ as a function of $T$. Their lengthy explicit expressions are not illuminating and hence not shown here. The first eigenvalue, $H_1$, coincides with the naive scalar meson (curvature) mass, $m_\sigma^2 =\Omega_{\sigma\sigma}$, in vacuum. It is nonzero, also at the CEP. The eigenvalues $H_2$ and $H_3$ are associated with $\Omega_{a_3 a_3}$ and $\Omega_{a_8 a_8}$ in vacuum. We find extended regions around the CEP where the real parts of these eigenvalues, shown in the left plot of \Fig{fig:EV}, are degenerate. In these regions they have imaginary parts with opposite signs but equal magnitude, see the right plot of \Fig{fig:EV}. This is an explicit realization of the possibility of the $\mathcal{CK}$-symmetric Hessian to have complex conjugate pairs of eigenvalues, $H_2 = H_3^*$.  
Thus, matter is in the complex regime discussed in \Sec{sec:mod}.

%%
%%%%%%%%%%
\begin{figure*}
\centering
\includegraphics[width=.325\textwidth]{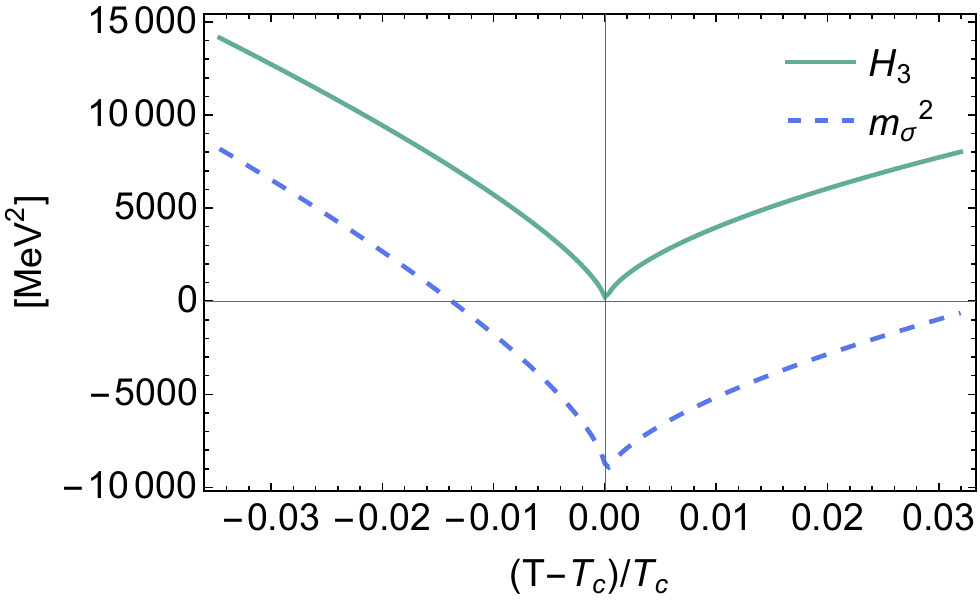}
\hfill
\includegraphics[width=.325\textwidth]{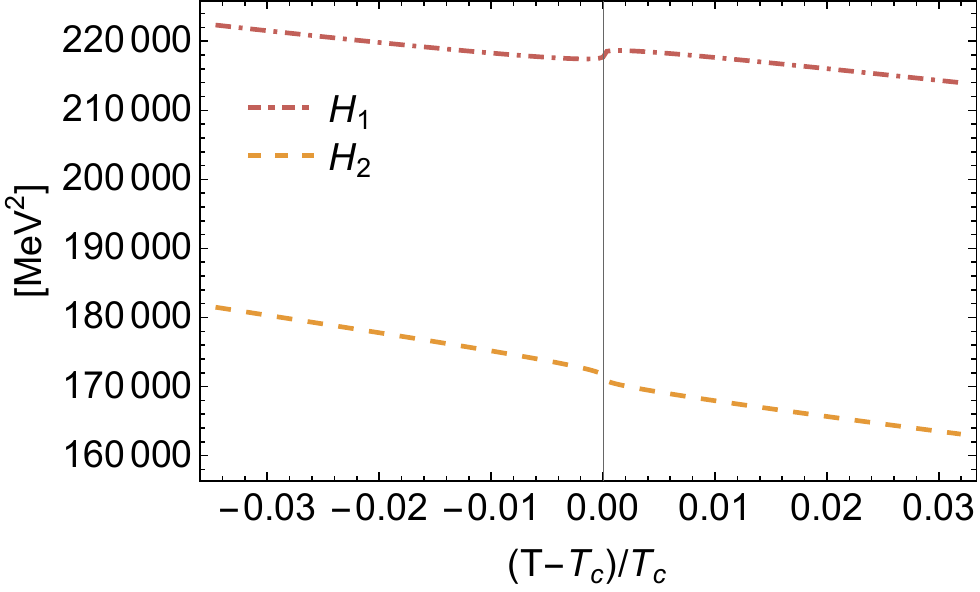}
\hfill
\includegraphics[width=.325\textwidth]{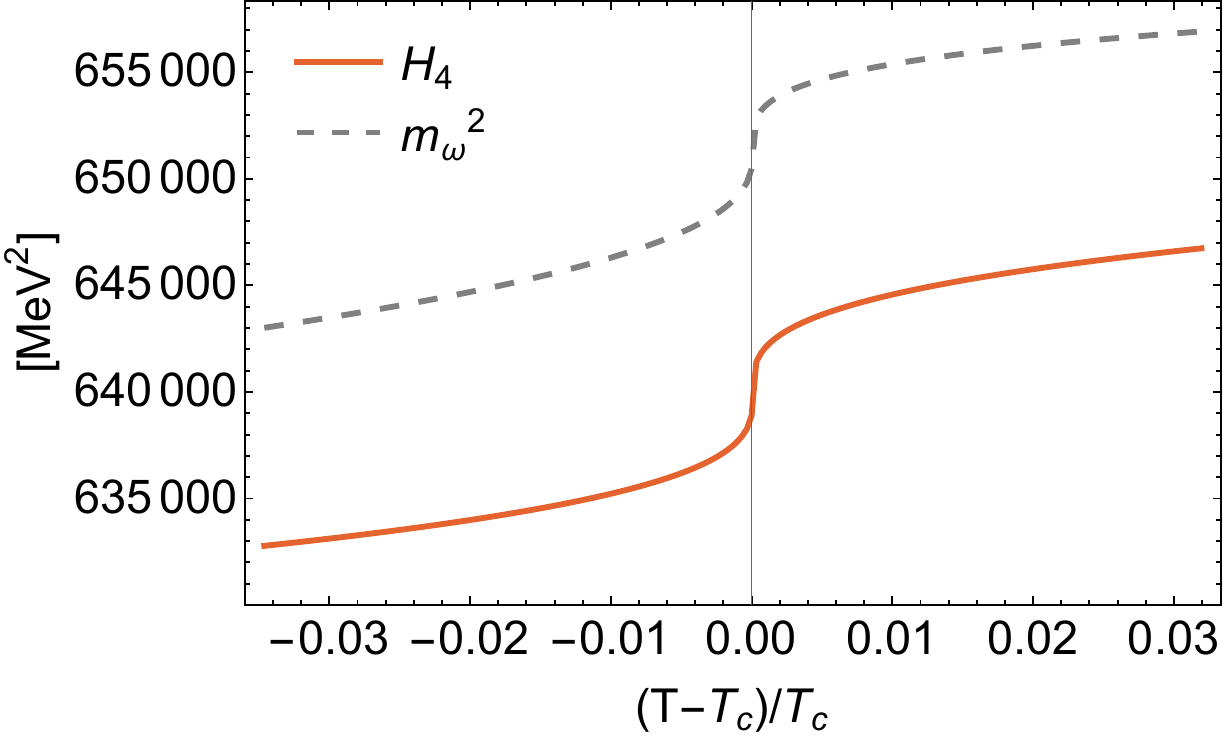}
\caption{Eigenvalues $H_1$, $H_2$, $H_3$ and $H_4$ of the Hessian of the ${\rm PQM_V}$ at the critical chemical potential as functions of $T$. The second eigenvalue is compared to the naive $\sigma$ meson mass (dashed blue line in the left plot) and the fourth eigenvalue to the naive $\omega$ meson mass (dashed gray line in the right plot). Without mixing, i.e.\ in vacuum, $H_1$ corresponds to the squared $\sigma$ meson mass, $H_4$ to the squared $\omega$ mass and $H_2$ and $H_3$ to temporal gluon correlations at zero momentum.}
\label{fig:EVPQMV}
\end{figure*}
%%%%%%%%%%
%%

Furthermore, $H_3$ shows a sharp drop at the CEP.  In order to see the behavior near the CEP more clearly, we zoom in on $H_3$ in its vicinity in \Fig{fig:massvsev}. We also show $m_\sigma^2$ in this region for comparison. Importantly, we observe that $m_\sigma^2$ does not vanish at the CEP. Thus, the naive scalar meson is not the critical mode. Its true nature is a mixture of the scalar meson and the Polyakov loops, or, loosely speaking, the eigenvalues of a temporal gluon background field. This is clearly demonstrated by the vanishing of $H_3$ at the CEP. We conclude that neither the curvature mass nor the screening mass of the sigma meson vanish at the CEP, but $\bar m_\chi^2 = H_3$ does. 

This confirms through an explicit example the general argument given in \Sec{sec:crit}: In the presence of mixing, the YLE manifests itself in the eigenvalues of the resulting Hessian. As a consequence, the critical mode must be a mixture of the involved fields/operators, which are the scalar meson and the Polyakov loops in the PQM. 

The intricate structure of the eigenvalues shown in \Fig{fig:EV} can be understood from an avoided crossing. To illustrate this, consider a non-Hermitian sub-matrix of the full Hessian, e.g.,
\begin{align}
    H^{(\rm sub)} = \begin{pmatrix} \Omega_{\sigma\sigma} & i \bar\Omega_{\sigma a_8} \\ i \bar\Omega_{\sigma a_8} & \Omega_{a_8 a_8}\end{pmatrix}\,,
\end{align}
where $\bar\Omega_{\sigma a_8}$ is real. The eigenvalues of $H^{(\rm sub)}$ are given by
\begin{align}
        H^{(\rm sub)}_{\pm} = \frac{\Omega_{\sigma\sigma}+\Omega_{a_8 a_8}}{2} \pm \frac{1}{2} \sqrt{\big(\Omega_{\sigma\sigma}-\Omega_{a_8 a_8}\big)^2-4\bar\Omega_{\sigma a_8}^2}\,.
\end{align}
From their difference,
\begin{align}
   H^{(\rm sub)}_{+}-H^{(\rm sub)}_{-} = \sqrt{\big(\Omega_{\sigma\sigma}-\Omega_{a_8 a_8}\big)^2-4\bar\Omega_{\sigma a_8}^2}\,,
\end{align}
it is obvious that either $H^{(\rm sub)}_{+}$ is always larger than $H^{(\rm sub)}_{-}$, or their real values degenerate and they get imaginary parts of equal magnitude and opposite sign. The latter is facilitated by the non-Hermitian Hessian. In any case, the real parts of the eigenvalues never cross. This is a generalization of the avoided crossing 
known from conventional quantum physics. We emphasize that this is a consequence of mixing. Eigenvalues of the mass matrix are allowed to cross if the corresponding modes do not mix, either because there is no mixing at all or because the Hessian becomes block-diagonal. However, it follows from our general analysis and we have also explicitly confirmed that at finite $T$ and $\mu$ all entries of the Hessian are nonzero.

Avoided crossing also explains why the critical mode, $H_3$, does not coincide with the pure $\sigma$-mode, i.e.\ eigenvalue that degenerates with $m_\sigma^2$ in the absence of mixing, $H_1$. We see from \Fig{fig:EV} that $H_1$ would have to cross the other eigenvalues in order to become critical. This is forbidden here. Note that this is specifically due to the mixing between the chiral condensate and the Polyakov loops.

It is also interesting to observe that the complex phase appears to be interrupted by the CEP. As seen in \Fig{fig:EV}, $H_3$ is complex right before and after the CEP, but real in its immediate vicinity.
The degeneracy of the real parts appears to be lifted exactly where the crossing of eigenvalues is avoided.
This indicates that a CEP and the complex phase are mutually exclusive. This follows from our discussion of \Sec{sec:crit}, since the edge singularity requires a vanishing eigenvalue, including its imaginary part. This is also intuitively clear. In the complex phase, the correlation functions associated with the complex eigenvalues show exponentially damped oscillatory behavior at large distances, cf.\ \Eq{eq:Gasymp}. The real part of the eigenvalue determines the inverse correlation length, while the wave number of the oscillation is given by the imaginary part. At the CEP the correlation length diverges. If this were to happen in the complex phase, the correlation function would be an oscillatory function with infinite range. This is characteristic for an inhomogeneous phase and therefore contradicts the assumption of homogeneous solutions of the EoM that has been made to find the CEP in the first place. So either an ordinary CEP exists without spatial modulations, or there are spatial modulations at the location of the CEP and its nature is changed, possibly to a Lifshitz point \cite{Hornreich:1975zz} or a Lifshitz regime \cite{Pisarski:2018bct}.

We emphasize that the fact that only one real eigenvalue vanishes here implies that the CEP still belongs to the universality class of the $3d$ Ising model. So while the nature of the critical mode of the QCD critical point is different than expected, its static universality is unchanged. We expect that this conclusion does not change in the realistic case, where the $\omega_0$ is added to the mix as well. This is veryfied next.

%%%%%%%%%%%%%%%%%%%%%%%%%%%%%%%%%%%%%%%%%%%%%%%%%%
%%%%%%%%%%%%%%%%%%%%%%%%%%%%%%%%%%%%%%%%%%%%%%%%%%
\subsection{Polyakov--Quark-Meson model with vector repulsion}\label{sec:PQMV}
%%%%%%%%%%%%%%%%%%%%%%%%%%%%%%%%%%%%%%%%%%%%%%%%%%
%%%%%%%%%%%%%%%%%%%%%%%%%%%%%%%%%%%%%%%%%%%%%%%%%%

As we have explicitly demonstrated in \Sec{sec:mix} there is also a medium-induced mixing to the isoscalar density $\omega_0$ in the vector channel. We therefore add a repulsive vector interaction to the PQM of the previous section for completeness. This is achieved by adding the following corrections to the effective action shown in \Eq{eq:SPQM},
\begin{align}\label{eq:DS}
    \begin{split}
        \Delta S &= \int_0^\beta\!dx_0 \int\!d^3x\Bigg\{ i h_\omega \bar\psi \gamma^0 \omega_0 \psi + \frac{1}{2} m_\omega^2 \omega_0^2
        \Bigg\}\,,
    \end{split}
\end{align}
so that the effective action of the PQM with a repulsive vector interaction (${\rm PQM_V}$) is
\begin{align}
    S_{\rm PQM_V} = S_{\rm PQM} + \Delta S\,.
\end{align}
This is the minimal extension of the PQM to capture the mean-field effects of $\omega_0$. It can be derived from microscopic interactions by performing a Hubbard-Stratonovich transformation of an effective, point-like four-quark interaction in the repulsive temporal vector channel $\sim (\bar\psi\gamma^0\psi)^2$\,. Of course, higher-order vector self-couplings and interactions between $\sigma$ and $\omega_0$ are also possible, but irrelevant for the present analysis. The thermodynamics of a similar model with an attractive $\omega_0$ interaction has been studied in \cite{Ueda:2013sia}.

%%
%%%%%%%%%%
\begin{figure*}[t]
\centering
\includegraphics[width=.46\textwidth]{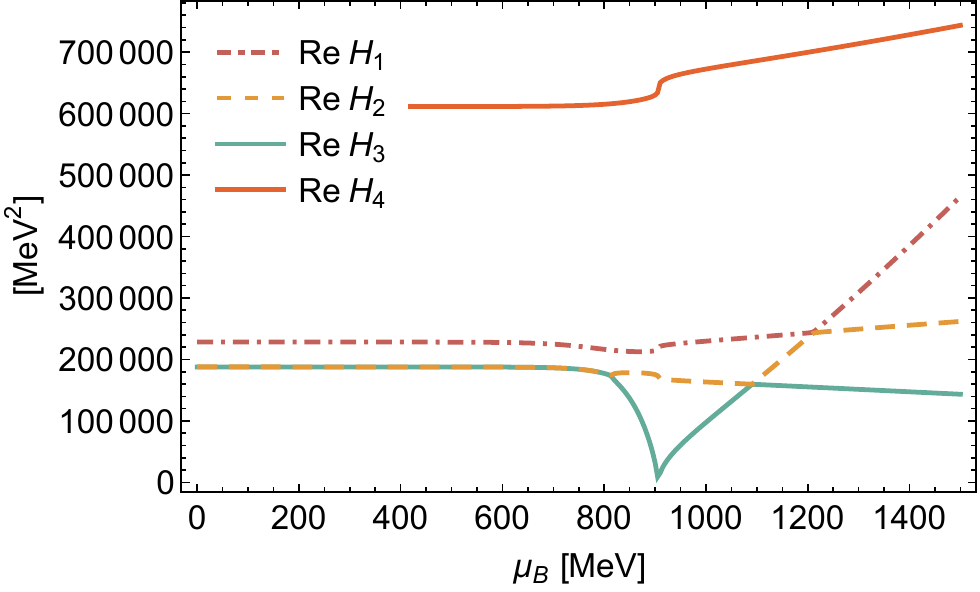}
\hfill
\includegraphics[width=.46\textwidth]{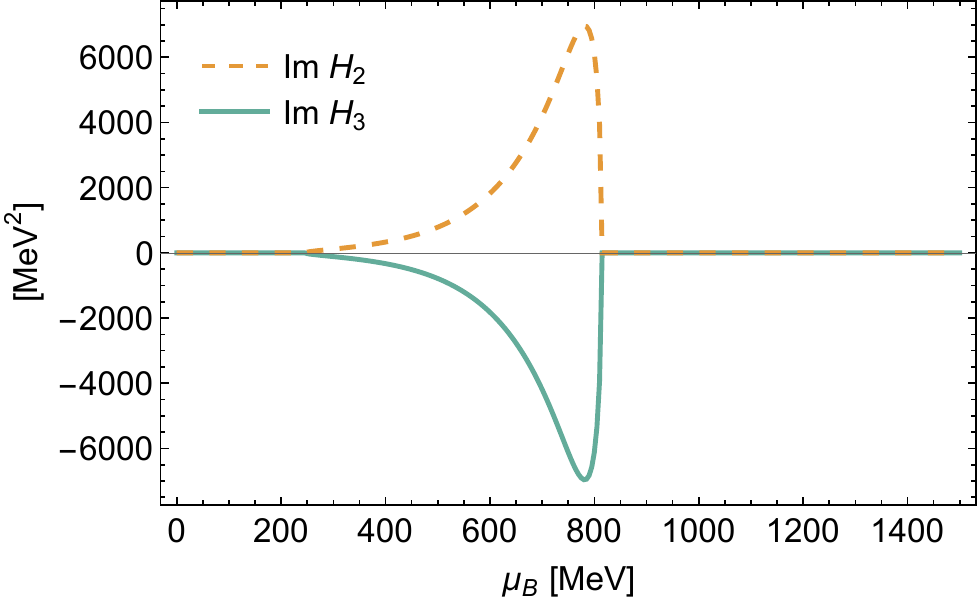}
\caption{Real (right plot) and imaginary (left plot) parts of the eigenvalues of the ${\rm PQM_V}$ at $T=T_{\rm CEP}$ as functions of baryon chemical potential. This shows the effects of avoided crossing on the eigenmodes and the presence of a complex regime.}
\label{fig:ImHPQMV}
\end{figure*}
%%%%%%%%%%
%%

Using that $\omega_0$ condenses at nonzero density, the resulting effective potential is
\begin{align}\label{eq:omega2}
\begin{split}
\Omega(\sigma,\omega_0,\Phi,\bar\Phi) &= V(\sigma) + \frac{1}{2} m_\omega^2 \omega_0^2 + U(\Phi,\bar\Phi)\\
&\quad- \frac{T}{\mathcal{V}}\Big[\ln\det\, \mathcal{M}\big(A_0,\sigma,\omega_0\big)\\
&\quad+ \ln\det\, \mathcal{M}_{\rm vac}\big(\sigma\big)\Big]\,.
\end{split}
\end{align}
The thermal part of the quark determinant is now identical to the one obtained from the leading order saddle point approximation in \Eq{eq:detM}. In the spirit of a more complete model, we also take the renormalized vacuum contribution of the quark determinant into account. Following, e.g., \cite{Skokov:2010sf, Mukherjee:2021tyg}, and using that the $\omega_0$ mean-field acts like a chemical potential, it reads after dimensional regularization and minimal subtraction of the divergent piece,
\begin{align}\label{eq:vacc}
\frac{T}{\mathcal{V}} \ln\det\, \mathcal{M}_{\rm vac}\big(\sigma\big) = N_f N_c \frac{h_\sigma^4\sigma^4}{2^8 \pi^2} \ln\Bigg(\frac{h_\sigma^2\sigma^2}{4 \Lambda^2}\Bigg)\,.
\end{align}
$\Lambda$ is the renormalization scale. This contribution renormalizes the quartic meson coupling $\lambda$ in \Eq{eq:VS} and as such affects the EoM of $\sigma$ and the scalar and pseudoscalar meson masses. This affects the parameters of the bosonic potential, for which we now choose $\nu = 107$\,MeV and $\lambda = 20$. This yields $f_\pi = 93.3$\,MeV and $m_\sigma = 477.7$\,MeV. In addition, we fix the $\omega$ mass parameter as commonly done to $m_\omega = 782$\,MeV and set the vector Yukawa coupling to $h_\omega = 1$. We note, however, that the only independent new parameter from the vector interaction is the ratio $m_\omega/h_\omega $, and that this parameter is not really related to the mass of the physical isoscalar vector meson. We refer to \cite{Jung:2019nnr} for a related discussion. 
All other parameters are unchanged from above. We furthermore note that the vacuum fluctuations move the CEP to significantly smaller $T$ and larger $\mu_B$. We counteract this effect to some extent by choosing a smaller $m_\sigma$ as compared to the previous section. The resulting CEP is then at
\begin{align}
(T_{\rm CEP}, \mu_{B,{\rm CEP}})\big|_{\rm PQM_V} \approx (60.1,906)\, {\rm MeV}\,.
\end{align}
The Hessian is a $4\!\times\! 4$ matrix,
\begin{align}\label{eq:HPQMV}
H_{\rm PQM_V} = \begin{pmatrix} \Omega_{\sigma\sigma}  &\Omega_{\sigma a_3} & \Omega_{\sigma a_8} & \Omega_{\sigma \omega_0} \\
\Omega_{\sigma a_3} & \Omega_{ a_3 a_3} & \Omega_{ a_3 a_8} & \Omega_{a_3 \omega_0} \\
\Omega_{\sigma a_8} & \Omega_{ a_3 a_8} & \Omega_{ a_8 a_8} & \Omega_{a_8 \omega_0} \\
\Omega_{\sigma \omega_0} & \Omega_{a_3 \omega_0} & \Omega_{a_8 \omega_0} & \Omega_{\omega_0 \omega_0}
 \end{pmatrix}\,.
\end{align}
Contributions of the quark determinant to its entries are given by the expressions in \Sec{sec:mix} plus a corresponding second derivative of the vacuum term in \Eq{eq:vacc}. The contributions from the purely bosonic part of the effective potential can be read off from Eqs.\ \eq{eq:VS}, \eq{eq:VP} and \eq{eq:DS}. Again, the conversion from Polyakov loops to $\bar A_0$ eigenvalues is done by Eqs.\ \eq{eq:PLpara} and \eq{eq:PLconv}.

We repeat the analysis of the eigenvalues of the Hessian near the CEP of this model as in the previous section. The results for the four eigenvalues of $H_{\rm PQM_V}$ at the critical chemical potential are shown in \Fig{fig:EVPQMV}. The eigenvalues $H_1$, $H_2$, $H_3$ and $H_4$ are associated with $\Omega_{\sigma\sigma}$, $\Omega_{a_3 a_3}$, $\Omega_{a_8 a_8}$ and $\Omega_{\omega_0\omega_0}$ in the absence of mixing. In the small region around the CEP shown here, all eigenvalues are real and there is no complex regime. 

As for the PQM in the previous section, we again find that the critical mode is associated with $H_3$, not as naively expected with $H_1$. Also here this is a result of avoided crossing. We compare $H_3$ to the $\sigma$ curvature mass in the left plot of \Fig{fig:EVPQMV}. Due to in-medium mixing, they are not the same. Analogous to the above, we find that the critical mode is a mixture of $\sigma$, $\omega_0$ and the Polyakov loops.

We can use the (in general complex) eigenvector $v_n$ to the eigenvalue $H_n$ of the Hessian to quantify this mixture. To this end, note that the Hessian is in the basis defined by $\phi = (\sigma,a_3,a_8,\omega_0)$. We can therefore use 
the normalized modulus of the components of the eigenvectors,
\begin{align}
    \hat v_n^i = \frac{|v_n^i|}{\sum_i |v_n^i|}\,,
\end{align}
to read-off the mixing from
\begin{align}
    \hat v_n^i \phi_i =\hat v_n^1\, \sigma + \hat v_n^2\, a_3 + \hat v_n^3\, a_8+ \hat v_n^4\, \omega_0 \,.
\end{align}
At the CEP we find
\begin{align}
  \hat v_3(T_{\rm CEP}, \mu_{B,{\rm CEP}})  \approx (0.807 , 0.082 , 0.003 , 0.108)\,,
\end{align}
which means that about 80\% of the critical mode stems from the chiral condensate and the rest from the density and the Polyakov loops in roughly equal parts. This remarkable given that the critical mode is associated to the eigenvalue $H_3$.

From the left plot of the left plot of \Fig{fig:EVPQMV} it is also apparent that the naive curvature mass $\Omega_{\sigma\sigma}$ is in general unphysical, as it becomes negative close to the CEP. Note that this does not entail that $\sigma$-like fluctuations are unstable, as these cannot be considered separately in the presence of mixing. A similar observation has been made in models for nuclear matter with a repulsive vector interaction \cite{Nishimura:2016yue}. We emphasize that the CEP is signaled by the diverging susceptibility in \Eq{eq:susc}. This is triggered by the vanishing of (at least) one eigenvalue of the Hessian. Some zero entry in the Hessian alone, e.g., $\Omega_{\sigma\sigma} = 0$ as seen in the left plot of \Fig{fig:EVPQMV}, is of no physical meaning in the presence of mixing.

We also compare the eigenvalue $H_4$ in the right plot to the naive $\omega$ curvature mass, $m_\omega^2 = \Omega_{\omega_0\omega_0}$. Around the CEP about 80\% of $H_4$ stems from the density and 11\% from the chiral condensate, explaining the difference between this eigenvalue and $m_\omega^2$.

Away from the CEP, also this model features a complex regime. This is demonstrated in \Fig{fig:ImHPQMV}, where we show the real and imaginary parts of the eigenvalues at $T=T_{\rm CEP}$ and various baryon chemical potentials. Avoided crossing is again clearly seen in the real parts. The real parts of $H_2$ and $H_3$ are degenerate between $\mu_B\approx 230 - 810$\,MeV, with opposite imaginary parts. In contrast to the PQM, the complex regime does not arise in the immediate vicinity of the CEP. However, the avoided crossing between $H_1$, $H_2$ and $H_3$ is what appears to take the system out of this regime also here. 

The exact location of the complex regime is certainly model-dependent. But, together with Refs.\ \cite{Nishimura:2014kla,Nishimura:2014rxa,Nishimura:2015lit,Nishimura:2016yue,Schindler:2019ugo,Schindler:2021otf}, our results provide further evidence that this is a generic feature of systems with $\mathcal{C}\mathcal{K}$ symmetric Hessians.

%%%%%%%%%%%%%%%%%%%%%%%%%%%%%%%%%%%%%%%%%%%%%%%%%%
%%%%%%%%%%%%%%%%%%%%%%%%%%%%%%%%%%%%%%%%%%%%%%%%%%
\section{Conclusion}\label{sec:conc}
%%%%%%%%%%%%%%%%%%%%%%%%%%%%%%%%%%%%%%%%%%%%%%%%%%
%%%%%%%%%%%%%%%%%%%%%%%%%%%%%%%%%%%%%%%%%%%%%%%%%%

We have shown that the breaking of charge conjugation symmetry in a dense medium leads to an intricate mixing that directly affects
qualitative features of QCD matter at and around the critical point.
Repulsive mixing between the chiral condensate, the density and the Polyakov loops gives rise to a non-Hermitian Hessian. This is induced in particular by a repulsive vector interaction and the lifted degeneracy between the expectation values of the Polyakov loop and its conjugate at finite density. However, the system retains a residual $\mathcal{C}\mathcal{K}$ symmetry and the Hessian can either have purely real eigenvalues or they come in complex conjugate pairs. The former occur in ordinary homogeneous phases, while the latter give rise to a patterned regime where correlations show spatial modulations as in a liquid.

Since the critical point is a Yang-Lee edge singularity, the critical mode can be identified by the eigenmodes to the vanishing eigenvalues of the static Hessian. It is hence a mixture of chiral condensate, the density and the Polyakov loops. In particular the admixture of the Polyakov loops, which are related to the free energy of single quarks, has been overlooked so far. Furthermore, the mixing leads to an avoided crossing of the eigenvalues of the Hessian. The critical mode is therefore in general not connected to the chiral condensate in vacuum if there are lower-lying eigenmodes in the system.

We found that a patterned regime with complex-conjugate eigenvalues and an ordinary CEP are mutually exclusive. If the critical mode would have spatially modulated correlations with infinite range, the system would be in an inhomogeneous phase. Formally, this follows from the requirement that both real and imaginary part of the eigenvalue of the Hessian vanish for the critical mode at the CEP.

We have studied the static eigenmodes near the CEP in different variations of the PQM model which exhibit these features in mean-field approximation. This allowed us to illustrate the physical consequences of medium-induced mixing using phenomenologically successful low-energy models of QCD. An exhaustive study of the parameter dependence and the full phase structure of these models in the presence of mixing is deferred to future work.
However, we emphasize that due to the substantial mixing, the complex regime in the phase diagram of these models seems to be a robust feature.

In addition to complex eigenvalues of the Hessian, spatial modulations can arise from other features of the self-energies of particles. For example, particles can be in a moat regime, where they have minimal energy at nonzero momentum \cite{Fu:2019hdw, Pisarski:2020dnx, Rennecke:2021ovl, Rennecke:2023xhc}. Or there can be an instability towards the formation of an inhomogeneous phase \cite{Buballa:2014tba}. Note that the latter is a special case of the moat regime, where the energy at the bottom of the moat vanishes. If and how the complex and the moat regime are connected is not entirely clear. It is known that a moat regime can arise even if mixing is not taken into account, see, e.g., Refs.\ \cite{Fu:2019hdw, Koenigstein:2021llr}. Hence, while a complex regime is sufficient for spatial modulations, it is not necessary. Understanding the relation and interplay between these different features is an interesting open problem. In any case, their ubiquity is a strong indication that spatially modulated regimes are natural in dense systems.

We emphasize that our analysis is not limit not the vicinity of the CEP of QCD at finite baryon chemical potential. It can be viewed as a blueprint to study the effects of mixing in various scenarios. Within QCD, additional mixing arises at finite strangeness and isospin chemical potentials as well as in the presence of other order parameters, such as diquarks in the color-superconducting phase at very large density. The nature of the critical modes at other critical points can also change, with potentially far reaching consequences for dynamical scaling.

%%%%%%%%%%%%%%%%%%%%%%%%%%%%%%%%%%%%%%%%%%%%%%%%%%
\begin{acknowledgments}
We are grateful to Laura Classen, Philippe de Forcrand, Milad Ghanbarpour, Theo Motta, Zohar Nussinov, Michael Ogilvie, Laurin Pannullo, Jan Pawlowski, Robert Pisarski, Stella Schindler and Marc Winstel for inspiring discussions and collaboration on related topics. M.H.\ acknowledges support by the DAAD student exchange program between the JLU Giessen and the University of Washington, ISAP-PHYSIK-JLU-UW(Seattle) 57575165. 
This work is supported by the Deutsche Forschungsgemeinschaft (DFG, German Research Foundation) through the Collaborative Research Center TransRegio CRC-TR 211 "Strong-interaction matter under extreme conditions"-- project number 315477589 -- TRR 211.
\end{acknowledgments}

%%%%%%%%%%%%%%%%%%%%%%%%%%%%%%%%%%%%%%%%%%%%%%%%%%
%%%%%%%%%%%%%%%%%%%%%%%%%%%%%%%%%%%%%%%%%%%%%%%%%%
%%%%%%%%%%%%%%%%%%%%%%%%%%%%%%%%%%%%%%%%%%%%%%%%%%
%%%%%%%%%%%%%%%%%%%%%%%%%%%%%%%%%%%%%%%%%%%%%%%%%%

\bibliography{crit}

\end{document}